\documentclass[9pt,shortpaper,twoside,web]{ieeecolor}
\usepackage{generic}
\usepackage{cite}
\usepackage{amsmath,amssymb,amsfonts}
\usepackage{algorithmic}
\usepackage{graphicx}
\usepackage{textcomp}
\usepackage{color}
\input{color.lrt}
\usepackage{tikz}
\usetikzlibrary{calc,shapes}
\usetikzlibrary{positioning,arrows}
\usepackage{pgfplots}
\usetikzlibrary{fit,calc}
\pgfplotsset{compat=1.14}
\usepackage{psfrag}
\usepackage{graphicx}
\usepackage{amsmath,amsfonts,amsxtra}
\usepackage{mathrsfs}
\usepackage{todonotes}
\usepackage{mathtools,leftidx}
\usepackage{float}
\usepackage{cleveref}
\crefname{equation}{}{}
\crefrangelabelformat{equation}{({#3}{#1}{#4}--{#5}{#2}{#6})}%
\crefrangelabelformat{subequation}{({#3}{#1}{#4}--{#5}\crefstripprefix{#1}{#2}#6)}%

\usepackage{environ}
\usepackage{bm} 
\allowdisplaybreaks[1]

\usepackage{caption}
\usepackage{subcaption}

\newcommand{\col}[1]{\operatorname{col}(#1)}
\newcommand{\cl}[1]{\operatorname{col}_1^{n_+}(#1)}

\newcommand{\diag}[1]{\operatorname{diag}(#1)}
\newcommand{\vect}[1]{\operatorname{vec}#1}

\renewcommand{\d}{\mathrm{d}}
\renewcommand{\t}{^{\top}}
\renewcommand{\c}{_{\text{c}}}
\renewcommand{\o}{_{\text{o}}}
\renewcommand{\cl}{_{\text{cl}}}
\newcommand{\tint}{\textstyle\int}

\newcommand{\e}{\text{e}}

\newcommand{\ltriag}[1][\textstyle]{
	\tikz[line join=round]{
		\node(a)[inner sep=0pt,outer sep=0pt]{\phantom{$#1 2$}};
		\draw(a.north west)--(a.south west)--(a.south east)--cycle;
	}
}

\newcommand{\utriag}[1][\textstyle]{
	\tikz[line join=round]{
		\node(a)[inner sep=0pt,outer sep=0pt]{\phantom{$#1 2$}};
		\draw(a.north west)--(a.south east)--(a.north east)--cycle;
	}
}

\newcommand{\N}{\bm{N}} 
 
\newcommand{\M}{\bm{M}}

\newcommand{\G}{\bm{\tilde{G}}}
 
\newcommand{\Si}{\bm{S}}

\newcommand{\B}{\bm{B}} 

\tikzset{
	state/.style={circle,draw,minimum size=6ex},
	arrow/.style={-latex, shorten >=1ex, shorten <=1ex}}

\newtheorem{df}{Definition}
\newtheorem{thm}{Theorem}
\newtheorem{Lemma}{Lemma}
\newtheorem{rem}{Remark}
\newtheorem{ass}{Assumption}

\def\BibTeX{{\rm B\kern-.05em{\sc i\kern-.025em b}\kern-.08em
    T\kern-.1667em\lower.7ex\hbox{E}\kern-.125emX}}
\begin{document}
\title{Backstepping Control of Coupled General Hyperbolic-Parabolic\\ PDE-PDE Systems}
\author{Joachim~Deutscher,~\IEEEmembership{Member,~IEEE,} Nicole~Gehring and Nick~Jung
\thanks{J. Deutscher and N. Jung are with the Institute of Measurement, Control and Microtechnology, Ulm University, Germany. (e-mail: {joachim.deutscher@uni-ulm.de; nick.jung@uni-ulm.de}).
N. Gehring is with the Institute of Automatic Control and Control Systems Technology, Johannes Kepler University Linz, Austria. (e-mail: nicole.gehring@jku.at)
}
}
\maketitle

\begin{abstract}
This paper considers the backstepping state feedback and observer design for hyperbolic and parabolic PDEs, which are bidirectionally interconnected in a general coupling structure. Both PDE subsystems consist of coupled scalar PDEs with the heterodirectional hyperbolic PDE subsystem subject to actuation and sensing. By making use of a multi-step approach to construct the transformation into a stable target system, it is shown that a backstepping state feedback and observer design only requires to solve the well-known kernel equations for the hyperbolic and parabolic subsystems as well as additional decoupling equations. The latter are standard initial boundary value problems for parabolic PDEs. This significantly facilitates the well-posedness analysis and the numerical computation of the backstepping controller. Exponential stability is verified for the state feedback loop, the observer error dynamics, and the closed-loop system using an observer-based compensator. The proposed backstepping design procedures are demonstrated for numerical examples.
\end{abstract}


\section{Introduction}
The backstepping control of distributed-parameter systems (DPSs) has attracted a lot of attention in the last two decades, with many contributions for  wide classes of PDEs (see \cite{Kr08,Vaz17}). The main idea of this approach is to use an inherently invertible Volterra integral transformation to map the DPS into a desired stable target system. 
Thereby, the main problem is to introduce a suitable target system so that the backstepping transformation exists. This is intimately related to the well-posedness of the kernel equations defining the backstepping transformation. Recent results in the literature consider the coupling of ODEs with PDEs resulting in so-called \emph{PDE-ODE systems}. In order to develop a systematic backstepping  design for them, two methods are available in the literature. The first one, called \emph{one-step approach} in the following, determines the backstepping transformation in one single step. Then, the resulting kernel equations consist of coupled PDEs and ODEs, which share a similar coupling structure with the underlying system to be controlled. Backstepping designs for bidirectionally coupled hyperbolic and parabolic PDE-ODE systems using this approach can be found in \cite{Di16,Wa19,Wa21}. An alternative method is the so-called \emph{multi-step approach} that exploits the coupling structure to construct the backstepping transformation in several successive steps. Thereby, the definition of each step is directly linked to the \emph{strict feedback form} of the underlying PDE-ODE system allowing a systematic determination of the required transformations and a simple successive stabilization (see \cite{Geh21}). Similarly, for systems in \emph{strict feedforward form}, the sequence of transformations for the observer design can be derived by making use of duality arguments. A major advantage of this approach is the fact that the resulting backstepping transformation is defined by the solution of already known and thus well-posed kernel equations as well as additional \emph{decoupling equations}. The latter appear due to the coupling between the ODE and PDE subsystems and have to be solved to ensure the decoupling of the subsystems. Consequently, verifying the existence of the backstepping mapping into a desired target system only requires to investigate the well-posedness of the decoupling equations. This problem can be easily solved, because the decoupling equations are initial value problems (IVPs) and boundary value problems for ODEs that take a simple form  in backstepping coordinates. This was demonstrated for hyperbolic and parabolic PDE-ODE systems in \cite{Deu18,Deu21a}. 

A current topic in backstepping control is its extension to bidirectionally coupled hyperbolic and parabolic PDEs with an actuation through the hyperbolic subsystem, which are called \emph{hyperbolic-parabolic PDE-PDE systems} in the sequel. This poses new problems in the design due to the different nature of the coupled subsystems, i.e., finite-time propagation versus instant diffusion. Furthermore, the actuation and sensing of only one of the PDEs gives rise to an underactuated control problem impeding the design (see also the results for underactuated hyperbolic PDE-PDE systems in \cite{Au20,Re22}). Besides the methodical interest in these systems, there also exists a strong background from the side of applications. In particular, they arise in biological chemotaxis, predator-prey models, extreme ultraviolet (EUV) lithography (see \cite{Chen17}) or chemical processes (see \cite[Ch. 11.2]{Kar19}). First backstepping results considered cascaded hyperbolic-parabolic PDE-PDE systems in \cite{Kr09c,Deu20}, which arise naturally from the modelling of input and output delays by transport equations. Recently, the much more challenging problem of the backstepping state feedback design for bidirectionally coupled scalar diffusion-reaction and transport equations has been dealt with in \cite{Chen17,Chen23} using the one-step approach. In these works, one has to verify the well-posedness of kernel equations consisting of coupled hyperbolic and parabolic PDEs, for which systematic solutions were presented. The extension of the backstepping approach to the most general case of hyperbolic-parabolic PDE-PDE systems, in which the subsystems themselves are coupled PDEs, has not been considered in the literature so far. 

In this paper, the multi-step approach is applied to the backstepping design of state feedback controllers and observers for a general class of hyperbolic-parabolic PDE-PDE systems. Thereby, these PDEs are bidirectionally interconnected by a general coupling structure and consist themselves of coupled scalar hyperbolic and parabolic PDEs. In particular, local, Volterra and Fredholm couplings are allowed in-domain, whereas local and integral couplings may appear at the boundaries. Different from the previous results in \cite{Chen17,Chen23}, a heterodirectional hyperbolic PDEs is considered instead of a single scalar transport equation giving rise to further challenges in the construction of the backstepping transformation. In particular, additional boundary couplings between the unactuated transport PDEs and the parabolic subsystem are taken into account, which lead to new challenges in the design. The actuation and sensing of the PDE-PDE system appears through the hyperbolic subsystem. In order to apply the backstepping design, general strict feedback and feedforward forms are identified for these systems. A major result of the paper is the fact that the multi-step approach allows to trace the design back to the well-known kernel equations for the hyperbolic and parabolic subsystems. Additionally, the resulting decoupling equations are a standard parabolic initial boundary value problem  (IBVP) so that its well-posedness can be verified using a modal approach. Consequently,  established numerical tools for solving kernel equations and standard solvers for parabolic PDEs are directly applicable to determine the backstepping state feedback and observer. These results are subsequently combined to obtain an observer-based compensator. 
Exponential stability of the resulting closed-loop state is verified.

The next section introduces the considered control problem. This is followed by the state feedback and observer design in Sections \ref{sec:sf} and \ref{sec:obsdes}, respectively. These results are combined in Section \ref{sec:obscomp} to determine the observer-based compensator. The last section illustrates the results of the paper using numerical examples. 



\emph{Notation.} For notational convenience, introduce the matrices 
\begin{equation}\label{Edef}
E_- = \begin{bsmallmatrix}
I_{n_{-}}\\
0
\end{bsmallmatrix} \in \mathbb{R}^{n \times n_-} , \quad 
E_+ = \begin{bsmallmatrix}
0\\
I_{n_+}
\end{bsmallmatrix} \in \mathbb{R}^{n \times n_+}.
\end{equation}
Then, the expressions $\leftidx{^-}{M} = E_-\t M$, $M_- = ME_-$, $\leftidx{^+}{M} = E_+\t M$ and $M_+ = ME_+$ can be defined for matrices $M$ of suitable dimensions. For vectors $v \in \mathbb{R}^n$, the notations $v_- = E_-\t v \in \mathbb{R}^{n_-}$ and $v_+ = E_+\t v \in \mathbb{R}^{n_+}$ are utilized in the sequel. 
The set of \emph{strictly lower triangular matrices} is denoted by $\ltriag(\mathbb{R}^{n \times n}) = \{M = [m_{ij}]\in \mathbb{R}^{n \times n}\,|\, m_{ij} = 0, i \leq j\}$. Similarly, the set of \emph{strictly upper triangular matrices} is given by $\utriag(\mathbb{R}^{n \times n}) = \{M = [m_{ij}]\in \mathbb{R}^{n \times n}\,|\, m_{ij} = 0, i \geq j\}$. The notations $\dot{x} = \partial_tx$, $x' = \partial_z x$ and $x'' = \partial_{zz} x$ are applied in the paper. For the parabolic subsystem, the weighted norm  $\|h\|_{\text{w}} = (\int_0^1\|\Theta^{-1/2}(\zeta)h(\zeta)\|^2_{\mathbb{C}^{n_-}}\d \zeta)^{1/2}$, $h(z) \in \mathbb{C}^{n_-}$, with the weight $\Theta(z)$ is used, while the norm $\|h\|_{l} = (\int_0^1\|h(\zeta)\|^2_{\mathbb{C}^{l}}\d \zeta)^{1/2}$, $h(z) \in \mathbb{C}^{l}$, $l \in \{n_-,n_+,n\}$, is utilized for the hyperbolic subsystem.

\section{Problem formulation}\label{sec:probform}
Consider the bidirectionally coupled \emph{hyperbolic-parabolic PDE-PDE system}
\begin{subequations}\label{plant}
	\begin{align}
	\dot{w} &= \Theta(z)w'' + \Phi(z)w' + F(z)w + \mathcal{H}_1[x](z) &&\label{parasys}\\
	w'(0) &= Q_0w(0) + \mathcal{H}_2[x], \; w(1) = \mathcal{H}_3[x]   &&\label{parasys3}\\
	\dot{x} &= \Lambda(z)x' + A(z)x + \mathcal{G}_1[w](z)  &&\label{hypsys}\\
	x_+(0) &= Q_+x_-(0) + \mathcal{G}_2[w]               &&\label{Bcp}\\
	x_-(1) &= Q_-x_+(1) + u + \mathcal{G}_3[w]           &&\label{hyprfref3}\\    
	y &= x_-(0)                                          &&\label{hypout}
	\end{align}
\end{subequations}
with the states $w(z,t) \in \mathbb{R}^{n_-}$ and  $x(z,t) \in \mathbb{R}^{n}$, $n = n_- + n_+$, defined on $[0,1]\times \mathbb{R}^+$, the input $u(t) \in \mathbb{R}^{n_-}$ and the anti-collocated measurement $y(t) \in \mathbb{R}^{n_-}$. Whenever convenient, the spatial argument $z$ and the time argument $t$ are dropped to increase readability. The diffusion coefficients of the parabolic subsystem \crefrange{parasys}{parasys3} have the property $\Theta = \diag{\theta_{1},\ldots,\theta_{n_-}} \in (C^2[0,1])^{n_- \times n_-}$
with $\theta_1(z) > \ldots > \theta_{n_-}(z) \geq  \underline{\theta} > 0$, $z \in [0,1]$. 
The advection is characterized by $\Phi = \diag{\Phi_1,\ldots,\Phi_{n_-}} \in (C^1[0,1])^{n_- \times n_-}$, the reaction by $F \in (C^1[0,1])^{n_- \times n_-}$ and the boundary condition (BC) at $z = 0$ by $Q_0 = \diag{q_{1},\ldots,q_{n_-}} \in \mathbb{R}^{n_- \times n_-}$. In what follows, $\Phi = 0$ is assumed, because the corresponding advection term can always be removed by a Hopf-Cole-type transformation (see, e.g., \cite{Deu17}).  The hyperbolic subsystem \crefrange{hypsys}{hypout} has the transport velocities contained in $\Lambda  = \diag{\lambda^-_1,\ldots,\lambda^-_{n_-},\lambda^+_{1},\ldots,\lambda^+_{n_+}} \in (C^1[0,1])^{n \times n}$ with $\lambda^-_1(z) > \ldots > \lambda^-_{n_-}(z) \geq \underline{\lambda} > 0 > \overline{\lambda} \geq \lambda^+_{1}(z) > \ldots > \lambda^+_{n_+}(z)$, $z \in [0,1]$. Hence, the state is given by $x = \col{x_-,x_+} \in \mathbb{R}^n$, where $x_-(z,t) \in \mathbb{R}^{n_-}$ describes the transport in the negative $z$-direction and $x_+(z,t)\in \mathbb{R}^{n_+}$ in the opposite direction, i.e., the hyperbolic subsystem is \emph{heterodirectional}. Define the transport times $D_- = \sum_{i = 1}^{n_-}\tint_0^1\d\zeta/\lambda^-_{i}(\zeta)$  and $D_+ = \tint_0^1\d\zeta/|\lambda^+_{1}(\zeta)|$ as well as $D = D_- + D_+$. 
Furthermore, $A = [a_{ij}] \in (C[0,1])^{n \times n}$ with $a_{ii} = 0$ 
holds, which can always be ensured by a Hopf-Cole-type transformation (see \cite{Hu15a}). The initial conditions (ICs) of the system are $w(z,0) = w_0(z) \in \mathbb{R}^{n_-}$ and $x(z,0) = x_0(z) \in \mathbb{R}^n$. 

The bidirectional interconnection between the subsystems is described by the formal \emph{integral operators}
\begin{equation}
 \mathcal{H}_i[x](z) \!=\! \tint_0^1\!\!H_i(z,\zeta)x(\zeta)\d\zeta,\;
 \mathcal{G}_i[w](z) \!=\! \tint_0^1\!\!G_i(z,\zeta)w(\zeta)\d\zeta\label{fop}
\end{equation}	 
for $i = 1,2,3$, in which  
\begin{subequations}\label{HGmat}
\begin{align}
 H_i(z,\zeta) \!&=\! \begin{cases}
 H^1_{1}(z)E\t_-\delta(\zeta) 
  \!+\! H^2_{1}(z)E\t_+\delta(\zeta\!-\!1)\\
   \!+\! H^3_{1}(z)\delta(\zeta\!-\!z) + H^4_{1}(z,\zeta)\sigma(z-\zeta)\\
   + H^5_{1}(z,\zeta), &\hspace{-0.3cm} i = 1\\
 H_i^1E\t_-\delta(\zeta) \!+\! H_i^2E\t_+\delta(\zeta-1) \!+\! H^3_{i}(\zeta), &\hspace{-0.3cm} i = 2,3
 \end{cases}\\
 G_i(z,\zeta) &= \begin{cases}
G^1_{1}(z)\delta(\zeta) + G^2_{1}(z)\delta'(\zeta-1) + G^3_{1}(z)\delta(\zeta-z)\\ + G^4_{1}(z,\zeta)\sigma(z-\zeta) + G^5_{1}(z,\zeta), &\hspace{-1.6cm} i = 1\\
 G_i^1\delta(\zeta) + G^2_{i}\delta'(\zeta-1) + G^3_{i}(\zeta), &\hspace{-1.6cm} i = 2,3.
 \end{cases}
\end{align}
\end{subequations}
All matrices $H_i^j$, $G_i^j$ appearing in $H_{i}(z,\zeta)$, $G_{i}(z,\zeta)$, $i =  1,2,3$, are in $L_2(0,1)$. Furthermore, $\sigma$ denotes the \emph{step function} and $\delta$ is the \emph{delta function}, where $\tint_0^1\delta'(\zeta-1)w(\zeta)\d\zeta = w'(1)$. With this, the integral operators \eqref{fop} describe local terms, Volterra- and Fredholm-type integral operators in-domain as well as couplings with integrals, $x_-(0)$,  $x_+(1)$, $w(0)$ and $w'(1)$ at the boundaries. 

In this paper, state feedbacks, observers and compensators are designed for \eqref{plant} using the \emph{backstepping method} (see, e.g., \cite{Kr08}). This, of course, requires stabilizability of the resulting state feedback loop and the observer error dynamics. For ODEs, it is well-known that a system representation in strict feedback and strict feedforward form ensures structural controllability and observability. A similar result is shown in \cite{Deu18,Deu21a} to hold for PDE-ODE systems by designing state feedbacks and observers that achieve arbitrary decay rates for the resulting closed-loop system and the observer error dynamics. Consequently, strict feedback and strict feedforward forms are also introduced for \eqref{plant} ensuring stabilizability of the system and the observer error dynamics. 
\begin{df}[Strict feedback form]\label{ass}
 If $\det H_3^1 \neq 0$, $H^2_3 = H_3^3 = 0$ and $H_i = 0$, $i = 1,2$, then \eqref{plant} is in \emph{strict feedback form}. 
\end{df}
If the system \eqref{plant} is in this form, then the parabolic subsystem is only affected by the hyperbolic $x_-$-system at $z=1$, while the hyperbolic subsystem is coupled with the parabolic subsystem at all boundaries and in-domain. Additionally, the hyperbolic PDE \emph{fully actuates} the parabolic PDE, i.e., the number $n_-$ of BCs at $z = 1$ in \eqref{parasys3} coincides with the linearly independent number of driving hyperbolic states implying $\det H_3^1 \neq 0$. 
In the case $\det H_3^1 = 0$, the proposed backstepping design may still be possible (cf. Remark \ref{rem:2} in Sec. \ref{sec:cdecpupl} for details). Otherwise, a singular $H_3^1$ can be dealt with by representing the parabolic subsystem in modal coordinates so that a modal stabilization with the methods in \cite{Ko75,Xu96} is possible. With this, a Robin BC at \eqref{parasys3} can be handled in the same way. 
\begin{df}[Strict feedforward form]\label{ass2}
 If $\det G_3^2 \neq 0$,  $G_3^1 = G_3^3 = 0$ and $G_i = 0$, $i = 1,2$, then  \eqref{plant} is in \emph{strict feedforward form}. 
\end{df}
Here, the hyperbolic subsystem is coupled with the parabolic subsystem via the BC at $z = 1$, while the parabolic subsystem is coupled with the hyperbolic subsystem at all boundaries and in-domain. Thereby, a  \emph{full sensing} $G_3^2w'(1)$ is required at the boundary \eqref{hyprfref3} meaning that $\det G_3^2 \neq 0$. With this, the boundary measurement of the hyperbolic subsystem contains the necessary information of the parabolic subsystem so that the observer error dynamics for the hyperbolic-parabolic PDE-PDE system is stabilizable.
Similar to the strict feedback form, the cases $\det G_3^2 = 0$, a general $G_3^2 \in \mathbb{R}^{n_- \times n_w}$ with an arbitrary number $n_w$ of parabolic PDEs and an integral coupling instead of $w'(1)$ at \eqref{hyprfref3} can also be dealt with using a modal approach (cf. Remark \ref{rem:2} in Sec. \ref{sec:cdecpupl} and \eqref{feqico}). 

These particular forms not only ensure a stabilization and estimation for \eqref{plant} with any desired decay rate, but are also well-suited to apply a systematic multi-step design. This is shown in the paper.

\begin{figure}[t]
	\centering
	\begin{tikzpicture}[node distance=4em, every state/.style={draw,minimum size=9mm}]
	\node [state, scale = 0.7] (S2) {$\tilde{x}_-$};
	\node [state, right=of S2, scale = 0.7] (S3) {$\tilde{w}$};
	\node [state, right=of S3, scale = 0.7, double] (S4) {$\tilde{x}_+$};
	\draw [arrow, bend left = 10, color=red, dashed]  (S3) to (S2);
	\draw [arrow, bend left = 10]  (S2) to (S3);
	\draw [arrow] (S3) to (S4);
	\draw [arrow, bend left = 20] (S2) to (S4);
	\end{tikzpicture}
	\caption{Coupling structure of \eqref{plantt} with the coupling (red dashed) to be eliminated and the $\tilde{x}_+$-subsystem stabilized (double ruled). The input $\tilde{u}$ is not shown.}\label{fig:sf1}%
\end{figure}
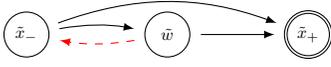
%


\section{State feedback design}\label{sec:sf}

\subsection{Controller Backstepping Transformation}\label{sec:bcoord}
For a system \eqref{plant} in \emph{strict feedback form} (see Definition \ref{ass}), the \emph{controller backstepping transformation}
\begin{subequations}\label{bcoord}
	\begin{align}
	\tilde{w} &= w - \tint_0^zP(z,\zeta)w(\zeta)\d\zeta = \mathcal{T}_w[w]\\ 
	\tilde{x} &= x - \tint_0^zK(z,\zeta)x(\zeta)\d\zeta = \mathcal{T}_x[x]\label{sftrafx}
	\end{align}	 
\end{subequations}
with the kernels $P(z,\zeta) \in \mathbb{R}^{n_- \times n_-}$, $K(z,\zeta) \in \mathbb{R}^{n \times n}$ and the new input
\begin{equation}\label{sff1}
\tilde{u} = \mathcal{G}_3[w]  + Q_-x_+(1) - \tint_0^1\leftidx{^-}{K}(1,\zeta)x(\zeta)\d\zeta + u
\end{equation}
are considered. Thereby, the kernels $P_I(z,\zeta)$ and $K_I(z,\zeta)$ of the inverse backstepping transformations $w = \tilde{w} + \int_0^zP_I(z,\zeta)\tilde{w}(\zeta)\d\zeta$, $x = \tilde{x} + \int_0^zK_I(z,\zeta)\tilde{x}(\zeta)\d\zeta$ directly follow from solving the corresponding reciprocity relations (see, e.g., \cite{Hu19}). Then, the resulting intermediate target system reads
\begin{subequations}\label{plantt}
	\begin{align}
	\dot{\tilde{w}} &= \Theta(z)\tilde{w}'' - \mu\c\tilde{w} + \tilde{F}_{0}(z)\tilde{w}(0)\label{parasyst}\\
	\tilde{w}'(0) &= 0, \;
	\tilde{w}(1) = -\tint_0^1P_I(1,\zeta)\tilde{w}(\zeta)\d\zeta + H_3^1\tilde{x}_-(0)\label{parasys3t}\\
	\dot{\tilde{x}} &= \Lambda(z)\tilde{x}' + \tilde{A}_0(z)\tilde{x}_-(0) + \tilde{\mathcal{G}}_1[\tilde{w}](z)\label{hypsyst}\\
	\tilde{x}_+(0) &= Q_+\tilde{x}_-(0) + \tilde{\mathcal{G}}_2[\tilde{w}],\;  \tilde{x}_-(1) = \tilde{u}\label{hyprfref3t}	
	\end{align}
\end{subequations}
with $\mu_c \in \mathbb{R}$, $\tilde{F}_0(z) \in \ltriag(\mathbb{R}^{n_- \times n_-})$, $\tilde{A}_0(z) = \col{\leftidx{^-}{\tilde{A}}_0(z),\linebreak \leftidx{^+}{\tilde{A}}_0(z)} \in \mathbb{R}^{n \times n_-}$, $\leftidx{^-}{\tilde{A}}_0(z) \in \ltriag(\mathbb{R}^{n_- \times n_-})$ and the operators  $\tilde{\mathcal{G}}_1$ and $\tilde{\mathcal{G}}_2$
given in the appendix. Hence, the transformation \eqref{sftrafx} decouples the $x_-$-system from the $x_+$-system and stabilizes the latter system (see Fig. \ref{fig:sf1}).

It can readily be verified that the kernels in \eqref{bcoord} have to satisfy the \emph{controller kernel equations} 
\begin{subequations}\label{app:k1}
	\begin{align}
	&\Theta P_{zz} - (P(z)\Theta)_{\zeta\zeta}(\zeta)  = P(F(\zeta) + \mu\c I)\\
	&(P(z)\Theta)_{\zeta}(0) = P(z,0)\Theta(0)Q_0 - \tilde{F}_0(z)\\ 
	&\Theta P'(z,z) + \Theta P_z(z,z) + (P(z)\Theta)_{\zeta}(z) = -(F(z) + \mu\c I)\\
	&P(z,z)\Theta(z) - \Theta(z)P(z,z) = 0, \; P(0,0) = Q_0\\
	&\Lambda K_{z} + (K(z)\Lambda)_{\zeta}(\zeta)  = KA(\zeta)\\
	&K(z,0)(\Lambda_-(0) + \Lambda_+(0)Q_+) = \tilde{A}_0(z)\\ 
	&K(z,z)\Lambda(z) - \Lambda(z)K(z,z) = A(z)
	\end{align}	 
\end{subequations}
defined on $0 \leq \zeta \leq z \leq 1$. It is verified in \cite{Vaz16a,Deu17,Hu15a,Hu19} that \eqref{app:k1} have a piecewise $C^2$-solution $P(z,\zeta)$ and a piecewise $C^1$-solution $K(z,\zeta)$.

\subsection{Controller Decoupling Transformation}\label{sec:cdecpupl}
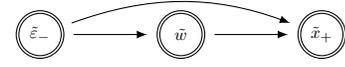
\begin{figure}[t]
	\centering
	\begin{tikzpicture}[node distance=4em, every state/.style={draw,minimum size=9mm}]
	\node [state, scale = 0.7, double] (S2) {$\tilde{\varepsilon}_-$};
	\node [state, right=of S2, scale = 0.7, double] (S3) {$\tilde{w}$};
	\node [state, right=of S3, scale = 0.7, double] (S4) {$\tilde{x}_+$};
	\draw [arrow]  (S2) to (S3);
	\draw [arrow] (S3) to (S4);
	\draw [arrow, bend left=20] (S2) to (S4);
	\end{tikzpicture}
	\caption{Coupling structure of the final target system \eqref{plantttf} with all subsystems stabilized (double ruled).}\label{fig:sf3}%
\end{figure}

In order to obtain a cascade for the final target system, the $\tilde{x}_-$-system has to be decoupled from the $\tilde{w}$-system (see Fig. \ref{fig:sf1}). 
This, however, leads to an unstable $\tilde{x}_-$-system, which has to be stabilized. Furthermore, the integral coupling in \eqref{parasys3t} has to be eliminated in order to obtain a stable parabolic subsystem. 
For this, introduce the \emph{controller decoupling} and \emph{backstepping transformation}
\begin{equation}
 \varepsilon_- \!=\! \tilde{x}_- \!-\! \tint_0^1\!\!N(z,\zeta)\tilde{w}(\zeta)\d\zeta,\;
 \tilde{\varepsilon}_- \!=\! \varepsilon_- \!-\!\tint_0^z\!\!Q(z,\zeta)\varepsilon_-(\zeta)\d\zeta\label{firstdecouplp}
\end{equation}	
with the kernels $N(z,\zeta), Q(z,\zeta) \in \mathbb{R}^{n_- \times n_-}$ mapping \eqref{plantt} into the final \emph{state feedback target system}
\begin{subequations}\label{plantttf}
	\begin{align}
	\dot{\tilde{\varepsilon}}_- &= \Lambda^-(z)\tilde{\varepsilon}'_- + \check{A}_0(z)\tilde{\varepsilon}_-(0)\label{hypsysttf}\\
	\tilde{\varepsilon}_-(1) &= 0\label{hyprfref3ttf}\\
	\dot{\tilde{w}} &= \Theta(z)\tilde{w}'' - \mu\c\tilde{w} + \tilde{F}_{0}(z)\tilde{w}(0)\label{parasysttf}\\
	\tilde{w}'(0) &= 0, \; \tilde{w}(1) = H_3^1\tilde{\varepsilon}_-(0)\label{parasys3ttf}\\
	\dot{\tilde{x}}_+ &= \Lambda^+(z)\tilde{x}'_+ + \leftidx{^+}{\tilde{A}}_0(z)\tilde{\varepsilon}_-(0) 
	+ \leftidx{^+}{\tilde{\mathcal{G}}}_1[\tilde{w}](z)\label{hypsysttf2}\\
	\tilde{x}_+(0) &= Q_+\tilde{\varepsilon}_-(0) + \tilde{\mathcal{G}}_2[\tilde{w}]\label{hypend}	              
	\end{align}
\end{subequations}
with $\Lambda^- = \diag{\lambda^-_1,\ldots,\lambda^-_{n_-}}$ and $\Lambda^+ = \diag{\lambda^+_1,\ldots,\lambda^+_{n_+}}$. 
Since $\check{A}_0(z)$ and  $\tilde{F}_{0}(z)$ are strictly lower triangular matrices, the  cascade of stable transport equations \crefrange{hypsysttf}{hyprfref3ttf} drives the cascade of stable parabolic PDEs \crefrange{parasysttf}{parasys3ttf} and the  parallel stable transport equations \crefrange{hypsysttf2}{hypend}, which is shown in Fig. \ref{fig:sf3}.

By differentiating \eqref{firstdecouplp} w.r.t. time, using \eqref{plantt} and \eqref{plantttf}, it can readily be shown that $N(z,\zeta)$ has to satisfy the \emph{controller decoupling equations}
\begin{subequations}\label{Ke:fs}
	\begin{align}
	\Lambda^-(z) N_{z} &= (N(z)\Theta)_{\zeta\zeta}(\zeta) - \mu\c N + \leftidx{^-}{R}(z,\zeta)\label{feqpde}\\
	(N(z)\Theta)_{\zeta}(0) &= \leftidx{^-}{\tilde{G}}_1^1(z) - \tint_0^1N(z,\zeta)\tilde{F}_0(\zeta)\d\zeta\label{feqrb1}\\
	N(z,1)\Theta(1) &= \leftidx{^-}{\tilde{G}}_1^2(z)\label{feqrb2}\\
	N(0,\zeta) &= (H_3^1)^{-1}P_I(1,\zeta)\label{feqic}
	\end{align}	 
\end{subequations}
defined on $(z,\zeta) \in [0,1]^2$ with $R(z,\zeta) = -\tilde{A}_{0}(z)(H_3^1)^{-1}P_I(1,\zeta) - G_1^3(z)\delta(\zeta-z) - \tilde{G}_1^4(z,\zeta)\sigma(z-\zeta) - \tilde{G}_1^5(z,\zeta)$. 
Therein, $z$ can be viewed as time and $\zeta$ as space variable so that \eqref{Ke:fs} is an IBVP for $N(z,\zeta) \in \mathbb{R}^{n_- \times n_-}$ with spatially and time varying coefficients. Consequently, \crefrange{feqrb1}{feqrb2} are the  BCs and \eqref{feqic} is the IC. Since a delta function appears in $R(z,\zeta)$, the decoupling equations \eqref{Ke:fs} have to be interpreted distributionally, i.e., in the weak sense. The next lemma asserts that \eqref{Ke:fs} still has a solution in $L_2$.
\begin{Lemma}[Controller decoupling equations]\label{lem:cfdecoupl}
	The controller decoupling equations \eqref{Ke:fs} have a unique mild solution $N(z,\cdot) \in (L_2(0,1))^{n_- \times n_-}$, $z \in [0,1]$ if $\mu\c$ is such that  \crefrange{parasysttf}{parasys3ttf} is exponentially stable.
\end{Lemma}
\begin{proof}
	In order to simplify the presentation, variables with double index $ij$ may be represented without the index but bold face, e.g., $\pmb{M} = M_{ij}$. Then, the elements $\N(z,\zeta) = N_{ij}(z,\zeta)$, $i, j = 1,\ldots,n_-$, of $N(z,\zeta)$ can be determined from \eqref{Ke:fs} by considering the \emph{component form}
	\begin{subequations}\label{cfredcomp}
		\begin{align}
		\lambda^{-}_i(z)\partial_z\N &= \partial_{\zeta\zeta}(\N(z)\theta_j)(\zeta) - \mu\c\N + \leftidx{^{\bm -}}{\bm{R}}(z,\zeta)\\
	    \partial_{\zeta}(\N(z)\theta_j)(0) &= \B_0(z),\;
		\N(z,1)\theta_j(1) = \leftidx{^{\bm -}}{\G}_1^2(z)\label{feqrb2p}\\
		\N(0,\zeta) &= \N_0(\zeta)
		\end{align}	 
	\end{subequations}
	with $\B_0(z) = \leftidx{^{\bm -}}{\G}_1^1(z) - \tint_0^1[N(z,\zeta)\tilde{F}_0(\zeta)]_{ij}\d\zeta$ and $\N_0(\zeta) = [(H_3^1)^{-1}P_I(1,\zeta)]_{ij}$.
	In view of $\tilde{F}_0(z) \in \ltriag(\mathbb{R}^{n_- \times n_-})$, the result
	\begin{equation}
	[N(z,\zeta)\tilde{F}_0(\zeta)]_{ij} \!=\! 
	\begin{cases}
	\sum\limits_{k = j+1}^{n_-}\!\!\!\!N_{ik}(z,\zeta)\tilde{F}_{0,kj}(\zeta), & \hspace{-0.2cm} 1 \!\leq\! j \!\leq\! n_-\!\!-\!\!1\\
	0,                                                                         & \hspace{-0.2cm} j = n_- 
	\end{cases}
	\end{equation}
	follows for $1 \leq i \leq n_-$, where $\tilde{F}_0(z) = [\tilde{F}_{0,ij}(z)]$. Hence, \eqref{cfredcomp} can be solved columnwise starting with the rightmost column and $\B_0$ representing a known inhomogeneity in \eqref{feqrb2p}. Introducing $\M(z,\zeta) = \N(z,\zeta)\theta_j(\zeta)$, the component form \eqref{cfredcomp} simplifies to
	\begin{subequations}\label{cfredcompt}
		\begin{align}
		\lambda^-_i(z)\partial_z\M &= \theta_j(\zeta)\partial_{\zeta\zeta}\M - \mu\c\M +  \Si(z,\zeta)\\
		\partial_{\zeta}\M(z,0) &= \Si_0(z),\;
		\M(z,1) = \leftidx{^{\bm -}}{\G}_1^2(z)\label{feqrb2pt}\\
		\M(0,\zeta) &= \M_0(\zeta),
		\end{align}	 
	\end{subequations}
	in which $\Si(z,\zeta) = \theta_j(\zeta)\leftidx{^{\bm -}}{\bm{R}}(z,\zeta)$, $\Si_0(z) =  \leftidx{^{\bm -}}{\G}_1^1(z) - \tint_0^1[M(z,\zeta)\tilde{F}_0(\zeta)]_{ij}/\theta_j(\zeta)\d\zeta$ and $\M_0(\zeta) = \theta_j(\zeta)\N_0(\zeta)$. By utilizing the change of coordinates
	\begin{equation}
	\M = \M^{\text{h}} + \underbrace{(\leftidx{^{\bm -}}{\G}_1^2(z) - \Si_0(z))\zeta^2 + \Si_0(z)\zeta}_{\M^{\text{ih}}},
	\end{equation}
	the component form
	\begin{subequations}\label{cfredcompt2}
		\begin{align}
		\lambda^-_i(z)\partial_z\M^{\text{h}} &= \theta_j(\zeta)\partial_{\zeta\zeta}\M^{\text{h}} - \mu\c\M^{\text{h}} +  \Si^{\text{h}}(z,\zeta)\\
		\partial_\zeta\M^{\text{h}}(z,0) &= \M^{\text{h}}(z,1) = 0\label{feqrb2pt2}\\
		\M^{\text{h}}(0,\zeta) &= \M^{\text{h}}_0(\zeta)
		\end{align}	 
	\end{subequations}
	subject to homogenous BCs is obtained with $\Si^{\text{h}}(z,\zeta) = \Si(z,\zeta) - \partial_z\M^{\text{ih}} + \theta_j(\zeta)\partial_{\zeta\zeta}\M^{\text{ih}} - \mu\c\M^{\text{ih}}$ and $\M^{\text{h}}_0(\zeta) = \M_0(\zeta) - \M^{\text{ih}}(0,\zeta)$. As $z$ plays the role of the time variable, \eqref{cfredcompt2} is an IBVP for a ``time-varying'' parabolic PDE.   In order to obtain a ``time-invariant'' PDE, consider the invertible change of variables $\tau = \phi_i^{-}(z) = \int_0^z(\d\zeta/\lambda^-_i(\zeta))$ and introduce $\tilde{\M}^{\text{h}}(\tau,\zeta) = \tilde{\M}^{\text{h}}(\phi_i^-(z),\zeta) = \M^{\text{h}}(z,\zeta)$. Then, simple calculations show that \eqref{cfredcompt2} is mapped into
	 \begin{subequations}\label{cfredcompt3}
	 	\begin{align}
	 	\partial_\tau\tilde{\M}^{\text{h}} &= \theta_j(\zeta)\partial_{\zeta\zeta}\tilde{\M}^{\text{h}}- \mu\c\tilde{\M}^{\text{h}} +  \tilde{\Si}^{\text{h}}(\tau,\zeta)\\
	 	\partial_{\zeta}\tilde{\M}^{\text{h}}(\tau,0) &= \tilde{\M}^{\text{h}}(\tau,1) = 0\label{feqrb2pt2a}\\
	 	\tilde{\M}^{\text{h}}(0,\zeta) &= \M^{\text{h}}_0(\zeta)
	 	\end{align}	 
	 \end{subequations}	
    on $\tau \in [0,\phi_i^-(1)]$, in which $\tilde{\Si}^{\text{h}}(\phi_i^-(z),\zeta) = \Si^{\text{h}}(z,\zeta)$. Define the operator $\mathcal{A}_jh = \theta_jh'' - \mu\c h$, $j = 1,\ldots,n_-$, $h \in D(\mathcal{A}_j) = \{h \in H^2(0,1)\;|\; h'(0) = h(1) =  0\} \subset X$ in the Hilbert space $X = L_2(0,1)$ endowed with the usual weighted inner product $\langle\cdot,\cdot\rangle$  inducing the norm $\|\cdot\|$. Since $-\mathcal{A}_j$ is a \emph{Sturm-Liouville operator}, there exists a sequence of eigenvectors $\{\phi^j_{k}, k \in \mathbb{N}\}$ w.r.t. the eigenvalues $\mu^j_{k}$ resulting from solving $\mathcal{A}_j\phi^j_k = \mu^j_k\phi^j_k$, $\phi^j_k \in D(\mathcal{A}_j)$, $k \in \mathbb{N}$, that are an (orthonormal) \emph{Riesz basis} for $X$ (see, e.g., \cite{Del03}). Consequently, $\tilde{\M}^{\text{h}}(\tau,\zeta) = \sum_{k=1}^{\infty}\tilde{\M}^{\text{h}}_k(\tau)\phi^j_k(\zeta)$, $\tilde{\M}^{\text{h}}_k(\tau) =\langle\tilde{\M}^{\text{h}}(\tau),\phi^j_k\rangle $ exists for any $\tilde{\M}^{\text{h}} \in X$. Substituting this in \eqref{cfredcompt3} and using $\M^{\text{h}}_{0,k} = \langle\M^{\text{h}}_0,\phi^j_k\rangle$, $\tilde{\Si}^{\text{h}}_{k}(\tau) = \langle\tilde{\Si}^{\text{h}}(\tau),\phi^j_k \rangle$, the IVP
	\begin{equation}\label{ivppc}
	\d_\tau\tilde{\M}^{\text{h}}_k(\tau) = \mu^j_k\tilde{\M}^{\text{h}}_k(\tau) + \tilde{\Si}^{\text{h}}_k(\tau), \quad k \in \mathbb{N},
	\end{equation}	 
	on $\tau \in [0,\phi_i^-(1)]$ with IC $\tilde{\M}_k^{\text{h}}(0) = \M^{\text{h}}_{0,k}$  results. Its solution is
	\begin{equation}\label{coeff}
	 \tilde{\M}^{\text{h}}_k(\tau) = \e^{\mu^j_k\tau}\M^{\text{h}}_{0,k} + \tint_0^\tau\e^{\mu^j_k(\tau-\bar{\tau})}\tilde{\Si}^{\text{h}}_k(\bar{\tau})\d\bar{\tau}.
	\end{equation} 
Note that $|\tilde{\Si}^{\text{h}}_k(\tau)|$  will not decay for $k \to \infty$ and $\tau \in [0,\phi_i^-(1)]$ if a delta function appears on the right-hand side of \eqref{feqpde}, but it is bounded. Nevertheless, convergence of the series for $\tilde{\M}^{\text{h}}(\tau,\zeta)$ can still be verified giving rise to an $L_2$-solution for \eqref{Ke:fs}. To this end, insert \eqref{coeff} in the corresponding series and use \emph{Parseval's equality} $\|\tilde{\M}^{\text{h}}(\tau)\|^2 = \sum_{k=1}^{\infty}|\tilde{\M}^{\text{h}}_k(\tau)|^2$, which is valid since $\{\phi^j_{k}, k \in \mathbb{N}\}$ is an orthonormal basis for $X$. Note that exponential stability of \crefrange{parasysttf}{parasys3ttf} implies $\max_{\lambda \in \sigma(\mathcal{A}_j)}\operatorname{Re}\lambda < 0$, yielding the estimate $\e^{2\mu^j_k\tau} \leq \e^{2\mu^j_{\text{max}}\tau}$, $k \in \mathbb{N}$, $\mu^j_{\text{max}} = \max_{k \in \mathbb{N}}\mu^j_k < 0$. With this, one obtains
\begin{equation}
 \|\tilde{\M}^{\text{h}}(\tau)\|^2 \leq \e^{2\mu^j_{\text{max}}\tau}\sum_{k=1}^{\infty}|\M^{\text{h}}_{0,k}|^2
 \!+ \bm{c}\sum_{k=1}^{\infty}\tint_0^\tau\! \e^{2\mu^j_k(\tau - \bar{\tau})}\d\bar{\tau},
\end{equation}	
in which $\bm{c} = \sup_{k \in \mathbb{N}, \tau \in [0,\phi_i^-(1)]}|\tilde{\Si}^{\text{h}}_k(\tau)|^2$. Then, 
the result
\begin{equation}
 \|\tilde{\M}^{\text{h}}(\tau)\|^2 \leq  \e^{2\mu^j_{\text{max}}\tau}\|\tilde{\M}^{\text{h}}(0)\|^2
+ \bm{c}\sum_{k=1}^{\infty}\tfrac{1}{2\mu^j_k}(\e^{2\mu^j_{\text{max}}\tau} - 1)
\end{equation}	
readily follows. Therein, the series converges because $|\mu^j_k| \in \mathcal{O}(k^2)$ holds for the Sturm-Liouville operator $-\mathcal{A}_j$ (see, e.g., \cite{Orl17a}). Hence, the series for $\tilde{\M}^{\text{h}}(\tau)$ converges in $L_2$ verifying that a unique solution exists in $L_2$ for $z > 0$, which equals the IC \eqref{feqic} for $z = 0$. This is the \emph{mild solution} of \eqref{cfredcompt3} and thus of \eqref{Ke:fs} (see, e.g., \cite[Def. 2.2.4]{Cu20}), which completes the proof.
\end{proof}
\begin{rem}
	Because of the inconsistent IC \eqref{feqic}, only a mild solution of the decoupling equations \eqref{Ke:fs} can be obtained. This means that the solution equals the IC \eqref{feqic} for $z = 0$, while the solution of the decoupling equations satisfies the BCs for $z > 0$. This, however, gives rise to a discontinuity in the solution at $z = 0$ due to the Dirichlet BC \eqref{feqic}. 	
	 Therefore, some care is required for the numerical solution of the decoupling equations \eqref{Ke:fs}. 
	 A discussion of this problem along with various methods for its solutions can be found in \cite{Boy99}. \hfill $\triangleleft$
\end{rem}%
\begin{rem}\label{rem:2}
In the general case, where $H_3^1 \in \mathbb{R}^{n_w \times n_-}$, i.e., an arbitrary number $n_w$ of parabolic PDEs with the state $w(z,t) \in \mathbb{R}^{n_w}$, the decoupling transformation \eqref{firstdecouplp} can still be utilized if $H_3^1N(0,\zeta) = P_I(1,\zeta)$, $\zeta \in [0,1]$, is solvable for $N(0,\zeta) \in \mathbb{R}^{n_- \times n_w}$ (cf. \eqref{feqic}). This may also be the case for $\det H_3^1 = 0$. Furthermore, if the system is \emph{overactuated}, i.e., $n_w < n_-$ and $\operatorname{rk}H_3^1 = n_w$, the decoupling equations \eqref{Ke:fs} are solvable. However, if the system is \emph{underactuated}, i.e., $n_w > n_-$, and no solution $N(0,\zeta)$ exists, then the modal approach mentioned in Section \ref{sec:probform} can be used.  \hfill  $\triangleleft$
\end{rem}
Furthermore, the kernel $Q(z,\zeta)$ in \eqref{firstdecouplp} has to satisfy the \emph{controller kernel equations}%
\begin{subequations}\label{app:k2}
	\begin{align}
	\Lambda^- Q_{z} + (Q(z)\Lambda^-)_{\zeta}(\zeta) &= 0, \; 0 < \zeta < z < 1\\
	Q(z,0)\Lambda^-(0) \!-\! \tint_0^zQ(z,\zeta)\Gamma_0(\zeta)\d\zeta &= \check{A}_0(z) \!-\!\Gamma_0(z)\\ 
	Q(z,z)\Lambda^-(z) - \Lambda^-(z)Q(z,z) &= 0,
	\end{align}	 
\end{subequations}
where $\Gamma_0(z) = \leftidx{^-}{\tilde{A}}_{0}(z) + ((N(z)\Theta)_\zeta(1) + \leftidx{^-}{\tilde{G}}_1^2(z) P_I(1,1))H_3^1$. By making use of the method of characteristics, it can be directly verified that \eqref{app:k2} has a piecewise $C^1$-solution $Q(z,\zeta)$ (see \cite{Deu18a}). The kernel $Q_I(z,\zeta)$ of the inverse backstepping transformation in \eqref{firstdecouplp} can readily be obtained from the reciprocity relation. With this, \eqref{firstdecouplp} is invertible, as the decoupling transformation can be directly solved for $\tilde{x}_-$. 
Then, the feedback
\begin{equation}\label{sfb2}
\tilde{u} = \tint_0^1N(1,\zeta)\tilde{w}(\zeta)\d\zeta +  \tint_0^1Q(1,\zeta)\varepsilon_-(\zeta)\d\zeta
\end{equation}
is obtained to get \eqref{plantttf}, which can be directly represented in terms of the original states $w$ and $x$ by inserting \eqref{bcoord}, \eqref{firstdecouplp}. 

The stability of the resulting closed-loop system with an assignable decay rate is stated in the next theorem.
\begin{thm}[State feedback controller]\label{zrstab}
Apply the state feedback controller \eqref{sff1}, \eqref{sfb2} to \eqref{plant}. Let $\mu_{\text{max}}$ be the largest eigenvalue of \crefrange{parasysttf}{parasys3ttf} for $\mu\c = 0$ and choose the design parameter $\mu\c$ such that $\alpha\c = \mu_{\text{max}} - \mu\c < 0$. 
Then, the closed-loop system is exponentially stable in the norm $\|\cdot\| = (\|\cdot\|_{\text{w}}^2 + \|\cdot\|_n^2)^{1/2}$, i.e., the closed-loop state $x_{\text{sf}}(t) = \{\col{w(z,t),x(z,t)}, z \in [0,1]\}$ satisfies
\begin{equation}
 \|x_{\text{sf}}(t)\| \leq M_{\text{c}}\e^{(\alpha\c + c)t}\|x_{\text{sf}}(0)\|, \quad t \geq 0,
\end{equation}%
for an $M_{\text{c}} \geq 1$, all ICs $w(\cdot,0) \in (L_2(0,1))^{n_-}$, piecewise differentiable ICs $x(z,0) \in \mathbb{R}^{n}$ and any $c > 0$ so that $\alpha\c + c < 0$. 	
\end{thm}	
\begin{proof}
Observe that \crefrange{hypsysttf}{parasys3ttf} is similar to the target system already investigated in the proof of Theorem 1 in \cite{Deu20}. While only parallel transport equations are considered in \cite{Deu20}, \crefrange{hypsysttf}{hyprfref3ttf} is a set of cascaded transport equations, which, however, does not change the result. Hence, the states $\tilde{\varepsilon}_-(t) = \{\tilde{\varepsilon}_-(z,t), z \in [0,1]\}$ and $\tilde{w}(t) = \{\tilde{w}(z,t), z \in [0,1]\}$ are piecewise continuous and thus bounded on $[0,D_-]$. With $\tilde{\varepsilon}_-(t) = 0$, $t > D_-$, the solution of the system \crefrange{parasysttf}{parasys3ttf} is exponentially convergent (see \cite{Deu20}), i.e., $\|\tilde{w}(t)\|_{\text{w}} \leq M_1\e^{(\alpha\c + c)(t-D_-)}\|\tilde{w}(D_-)\|_{\text{w}}$, $t > D_-$, for an $M_1 \geq 1$ and all ICs $\tilde{w}(\cdot,0) \in (L_2(0,1))^{n_-}$. Consequently, the state $\tilde{x}_+(t) = \{\tilde{x}_+(z,t), z \in [0,1]\}$ is bounded on $[0,D_-]$, because it can be verified with the same reasoning as in \cite{Deu20} that the operators $\leftidx{^+}{\tilde{\mathcal{G}}}_1$ and $\tilde{\mathcal{G}}_2$ are relatively bounded. This, in particular, leads to bounded operators when applied to the analytic $C_0$-semigroup generated by the parabolic subsystem.  Since \crefrange{hypsysttf2}{hypend} are a set of parallel transport equations, one can make use of the results in the proof of Theorem 2 of \cite{Deu20} to show that the solution of the system \crefrange{hypsysttf2}{hypend} is exponentially convergent for $t > D$, i.e.,  $\|\tilde{x}_+(t)\|_{n_+} \leq M_2\e^{(\alpha\c + c) (t-D)}\|\tilde{w}(D)\|_{\text{w}} \leq M_1M_2\e^{(\alpha\c + c) (t-D_-)}\|\tilde{w}(D_-)\|_{\text{w}}$, $t > D$, for an $M_2 > 0$ and all piecewise differentiable ICs $\tilde{x}_+(z,0) \in \mathbb{R}^{n_+}$. This implies $\|\col{\tilde{w}(t),\tilde{\varepsilon}_-(t),\tilde{x}_+(t)}\| = (\|\tilde{w}(t)\|^2_{\text{w}} + \|\tilde{\varepsilon}_-(t)\|^2_{n_-} + \|\tilde{x}_+(t)\|^2_{n_+})^{1/2} \leq (\max(M^2_1,(M_1M_2)^2))^{1/2}\linebreak\cdot\e^{(\alpha\c + c) (t-D_-)}\|\tilde{w}(D_-)\|_{\text{w}}$, $t > D$, for all ICs $\tilde{w}(\cdot,0)\linebreak \in (L_2(0,1))^{n_-}$ and all piecewise differentiable ICs $\tilde{\varepsilon}_-(z,0) \in \mathbb{R}^{n_-}$, $\tilde{x}_+(z,0) \in \mathbb{R}^{n_+}$. Since the target system state is bounded in $t \in [0,D]$ and the finite-time stable transport equations are exponentially stable in the $L_2$-norm for any decay rate (see, e.g., \cite{Hu19}), there exists a sufficiently large $M_3 > (\max(M^2_1,(M_1M_2)^2))^{1/2}$ so that the target system \eqref{plantttf} is exponentially stable for $t \geq 0$. Then, going through the chain of the boundedly invertible transformations, exponential stability in the original coordinates can be verified. 
\end{proof}	
\section{Observer design}\label{sec:obsdes}
For the system \eqref{plant} in \emph{strict feedforward form} (see Definition \ref{ass2}), the \emph{observer}
\begin{subequations}\label{obs}
	\begin{align}
	\dot{\hat{w}} &= \Theta(z)\hat{w}'' \!+\! F(z)\hat{w} \!+\! \mathcal{H}_1[\hat{x}](z) \!+\! L_w(z)(y \!-\! \hat{x}_-(0))\label{parasyso}\\
	\hat{w}'(0) &= Q_0\hat{w}(0) + \mathcal{H}_2[\hat{x}] + L_0(y - \hat{x}_-(0))\\
	\hat{w}(1) &=  \mathcal{H}_3[\hat{x}] + L_1(y - \hat{x}_-(0)) \label{parasys3o}\\
	\dot{\hat{x}} &= \Lambda(z)\hat{x}' + A(z)\hat{x} + L(z)(y - \hat{x}_-(0))\label{hypsyso}\\
	\hat{x}_+(0) &= Q_+y, \; \hat{x}_-(1) = Q_-\hat{x}_+(1)  + G_3^2\hat{w}'(1) + u\label{hyprfref3o}
	\end{align}
\end{subequations}
defined on $(z,t) \in [0,1]\times \mathbb{R}_0^+$ with ICs $\hat{w}(z,0) = \hat{w}_0(z) \in \mathbb{R}^{n_-}$ and $\hat{x}(z,0) = \hat{x}_0(z) \in \mathbb{R}^{n}$ is designed, in which $\hat{w}$ and $\hat{x}$ are estimates for the states $w$ and $x$ of the coupled hyperbolic-parabolic system \eqref{plant}. The observer gains $L_w(z), L_0, L_1 \in \mathbb{R}^{n_- \times n_-}$ and $L(z) \in \mathbb{R}^{n \times n_-}$ are determined to stabilize the corresponding \emph{observer error dynamics}%
\begin{subequations}\label{obse}
	\begin{align}
	\dot{e}_w &= \Theta(z)e_{w}'' + F(z)e_w  + \mathcal{H}_1[e](z) -L_w(z)e_-(0)\label{parasysoe}\\
	  e_w'(0) &= Q_0e_{w}(0) + \mathcal{H}_2[e] - L_0e_-(0) \\
	 e_{w}(1) &= \mathcal{H}_3[e] - L_1e_-(0) \label{parasys3oe}\\
	  \dot{e} &= \Lambda(z)e' + A(z)e - L(z)e_-(0)\label{hypsysoe}\\
	   e_+(0) &= 0, \;  e_-(1) = Q_-e_+(1) + G_3^2e'_w(1)\label{hyprfref3oe}
	\end{align}
\end{subequations}
with $e_w = w - \hat{w}$ and $e = x -\hat{x}$.

\subsection{Observer Backstepping Transformation}\label{sec:obsbc}
\begin{figure}[t]
	\centering
	\begin{tikzpicture}[node distance=4em, every state/.style={draw,minimum size=9mm}]
	\node [state, scale = 0.7] (S2) {$\tilde{e}_-$};
	\node [state, right=of S2, scale = 0.7] (S3) {$\tilde{e}_w$};
	\node [state, right=of S3, scale = 0.7, double] (S4) {$\tilde{e}_+$};
	\draw [arrow, bend left = 10]  (S3) to (S2);
	\draw [arrow, bend left = 10, color=red, dashed]  (S2) to (S3);
	\draw [arrow] (S4) to (S3);
	\draw [arrow, bend left = 20] (S4) to (S2);
	\end{tikzpicture}
	\caption{Coupling structure of \eqref{obset2} with the coupling (red dashed) to be eliminated and the $\tilde{e}_+$-subsystem stabilized (double ruled). The output injections are not shown.}\label{fig:obs1}%
\end{figure}
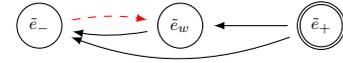

Similar to the state feedback design, the \emph{observer backstepping transformation}
\begin{subequations}\label{bcoordobs}
\begin{align}
 e_w &= \tilde{e}_w - \tint_z^1\bar{P}(z,\zeta)\tilde{e}_w(\zeta)\d\zeta = \bar{\mathcal{T}}^{-1}_w[\tilde{e}_w]\\
   e &= \tilde{e} - \tint_0^z\bar{K}(z,\zeta)\tilde{e}(\zeta)\d\zeta = \bar{\mathcal{T}}^{-1}_x[\tilde{e}]\label{btrafobs}
\end{align}	 	 
\end{subequations}
with the kernels $\bar{P}(z,\zeta) \in \mathbb{R}^{n_- \times n_-}$, $\bar{K}(z,\zeta) \in \mathbb{R}^{n \times n}$ is introduced to map \eqref{obse} into
\begin{subequations}\label{obset2}
\begin{align}
\dot{\tilde{e}}_w &= \Theta(z)\tilde{e}_{w}'' - \mu\o\tilde{e}_w + \bar{\mathcal{H}}_1[\tilde{e}](z)                
                     -M_w(z)\tilde{e}'_{w}(1)\nonumber\\
                  &\quad    \!-\!\tilde{L}_w(z)\tilde{e}_-(0)\label{parasysoet}\\
\tilde{e}_w'(0) &= -\tint_0^1\bar{F}_{0}(\zeta)\tilde{e}_w(\zeta)\d\zeta + \bar{\mathcal{H}}_2[\tilde{e}]\\
\tilde{e}_{w}(1) &= \bar{\mathcal{H}}_3[\tilde{e}]  \label{parasys3ot}\\
\dot{\tilde{e}} &= \Lambda(z)\tilde{e}' - \tilde{L}(z)\tilde{e}_-(0)\label{hypsysot}\\
\tilde{e}_+(0) &= 0 \\
\tilde{e}_-(1) &= Q_-\tilde{e}_+(1)  - \tint_0^1\bar{A}_{0}(\zeta)\tilde{e}(\zeta)\d\zeta\nonumber\\
              & \quad + G_3^2\bar{P}(1,1)\tilde{e}_w(1) + G_3^2\tilde{e}'_w(1),\label{hyprfref3ot}
\end{align}
\end{subequations}
in which $\mu\o \in \mathbb{R}$, $\bar{F}_{0}(z) \in \utriag(\mathbb{R}^{n_- \times n_-})$, $\bar{A}_0(z) = [\bar{A}_{0,-}(z)\;\; \bar{A}_{0,+}(z)] \in \mathbb{R}^{n_- \times n}$, $\bar{A}_{0,-}(z) \in \utriag(\mathbb{R}^{n_- \times n_-})$ and the operators  $\bar{\mathcal{H}}_1$, $\bar{\mathcal{H}}_2$ and $\bar{\mathcal{H}}_3$ given in the appendix. Obviously, the backstepping transformation \eqref{btrafobs} decouples the $e_+$-system from the $e_-$-system and stabilizes the former (see 
Fig. \ref{fig:obs1}). With the results in \cite {Vaz16a,Deu17,Hu15a,Hu19}, it can be verified that the kernels in \eqref{bcoordobs} satisfy the \emph{observer kernel equations}
\begin{subequations}\label{app:k1o}
	\begin{align}
	& \Theta\bar{P}_{zz} - (\bar{P}(z)\Theta)_{\zeta\zeta}(\zeta)  = -(F(z) + \mu\o I)\bar{P}\label{app:o1}\\
	&\bar{P}_{z}(0,\zeta) = Q_0\bar{P}(0,\zeta) - \bar{F}_{0}(\zeta)\\ 
	&\Theta \bar{P}'(z,z) + \Theta \bar{P}_z(z,z) + (\bar{P}(z)\Theta)_{\zeta}(z) = -(F(z) + \mu\o I)\\
	&\bar{P}(z,z)\Theta(z) - \Theta(z)\bar{P}(z,z) = 0, \; \bar{P}(0,0) = Q_0\label{app:o2}\\
	&\Lambda \bar{K}_{z} + (\bar{K}(z)\Lambda)_{\zeta}(\zeta)  = -A(z)\bar{K}\label{app:o3}\\
	&\leftidx{^-}{\bar{K}}(1,\zeta) - Q_-\leftidx{^+}{\bar{K}}(1,\zeta)  = - \bar{A}_0(\zeta)\\ 
	&\bar{K}(z,z)\Lambda(z) - \Lambda(z)\bar{K}(z,z) = -A(z)\label{app:o4}	
	\end{align}	 
\end{subequations}
with \crefrange{app:o1}{app:o2} defined on $0 \leq z \leq \zeta \leq 1$ and \crefrange{app:o3}{app:o4} on $0 \leq \zeta \leq z \leq 1$. They have a piecewise $C^2$-solution $\bar{P}(z,\zeta)$ and a piecewise $C^1$-solution $\bar{K}(z,\zeta)$, which is verified in 
\cite{Vaz16a,Deu17,Hu15a,Hu19}. The kernels $\bar{P}_I(z,\zeta)$ and $\bar{K}_I(z,\zeta)$ of the inverse backstepping transformation
$\tilde{e}_w = e_w +\! \int_z^1\!\bar{P}_I(z,\zeta)e_w(\zeta)\d\zeta$, $\tilde{e} = e +\! \int_0^z\!\bar{K}_I(z,\zeta)e(\zeta)\d\zeta$ can be determined using the corresponding reciprocity relations. 
The observer gains read
\begin{subequations}\label{obsgain1}
\begin{align}
 L(z)   &= \bar{\mathcal{T}}^{-1}_x[\tilde{L}](z) - \bar{K}(z,0)\Lambda_-(0)\label{obsgain1a}\\
 L_w(z) &= \bar{\mathcal{T}}^{-1}_w[\tilde{L}_w](z),\;
 M_w(z) = -\bar{\mathcal{T}}_w[\bar{P}(\cdot,1)\Theta(1)](z)\label{obsgain1b}\\
 L_0 &= H_2^1,\; L_1 = H_3^1.
\end{align}%
\end{subequations}%
\subsection{Observer Decoupling Transformation}\label{sec:obsfred}
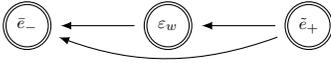
\begin{figure}[t]
	\centering
	\begin{tikzpicture}[node distance=4em, every state/.style={draw,minimum size=9mm}]
	\node [state, scale = 0.7, double] (S2) {$\bar{e}_-$};
	\node [state, right=of S2, scale = 0.7, double] (S3) {$\varepsilon_w$};
	\node [state, right=of S3, scale = 0.7, double] (S4) {$\tilde{e}_+$};
	\draw [arrow]  (S3) to (S2);
	\draw [arrow] (S4) to (S3);
	\draw [arrow, bend left = 20] (S4) to (S2);
	\end{tikzpicture}
	\caption{Coupling structure of the final target system \eqref{obset3} with  all subsystems stabilized (double ruled).}\label{fig:obs2}%
\end{figure}

A cascade structure is obtained by eliminating the red dashed coupling in Fig. \ref{fig:obs1}.
For this, the \emph{observer decoupling} and \emph{backstepping transformation}
\begin{equation}
 \varepsilon_w \!=\! \tilde{e}_w \!-\! \tint_0^1\!\!\bar{N}(z,\zeta)\tilde{e}_-(\zeta)\d\zeta,\;
     \tilde{e}_- \!=\! \bar{e}_- \!-\! \tint_0^z\!\!\bar{Q}(z,\zeta)\bar{e}_-(\zeta)\d\zeta\label{obsfinaltraf}
\end{equation}	
with the kernels $\bar{N}(z,\zeta), \bar{Q}(z,\zeta) \in \mathbb{R}^{n_- \times n_-}$ is considered. The first one decouples the $\tilde{e}_w$-system from the $\tilde{e}_-$-system and removes the local term $\tilde{e}_w'(1)$ in \eqref{parasysoet} resulting in a stable parabolic error dynamics. This, however, leads to an unstable $\tilde{e}_-$-system, which has to be stabilized by the backstepping transformation in \eqref{obsfinaltraf}. As a result, the final \emph{observer error target system}
\begin{subequations}\label{obset3}
\begin{align}
	\dot{\tilde{e}}_+ &= \Lambda^+(z)\tilde{e}_+'\label{hypsysot33}\\
	   \tilde{e}_+(0) &= 0\label{hyprfref3ot3}\\
	\dot{\varepsilon}_w &= \Theta(z)\varepsilon_{w}'' - \mu\o\varepsilon_w + \bar{\mathcal{H}}^+_1[\tilde{e}_+](z) \label{parasysoet3}\\
	 \varepsilon_w'(0) &= -\tint_0^1\!\bar{F}_{0}(\zeta)\varepsilon_w(\zeta)\d\zeta + \bar{\mathcal{H}}^+_2[\tilde{e}_+], \; \varepsilon_{w}(1)\! = \!\bar{\mathcal{H}}^+_3[\tilde{e}_+] \label{obsprb}\\
	\dot{\bar{e}}_- &= \Lambda^-(z)\bar{e}'_-\label{hypsysot3}\\
	   \bar{e}_-(1) &= Q_-\tilde{e}_+(1)   \!-\! \tint_0^1\hat{A}_{0}(\zeta)\bar{e}_-(\zeta)\d\zeta \!-\! \tint_0^1\bar{A}_{0,+}(\zeta)\tilde{e}_+(\zeta)\d\zeta\nonumber\\
	                  &\quad + G_3^2\bar{P}(1,1)\varepsilon_w(1) + G_3^2\varepsilon'_w(1) \label{obsfinendrb}
\end{align}
\end{subequations} 
is obtained. Therein, the operators $\bar{\mathcal{H}}^+_i[\tilde{e}_+] = \bar{\mathcal{H}}_i[E_+\tilde{e}_+]$, $i = 2,3$, and $\bar{\mathcal{H}}^+_1[\tilde{e}_+](z) = \bar{\mathcal{H}}_1[E_+\tilde{e}_+](z) - \bar{N}(z,1)\Lambda^-(1)(Q_-\tilde{e}_+(1) - \tint_0^1\bar{A}_{0,+}(\zeta)\tilde{e}_+(\zeta)\d\zeta + \bar{P}(1,1)\bar{\mathcal{H}}_3[E_+\tilde{e}_+])$ are utilized. Since $\bar{F}_{0}(z) \in \utriag(\mathbb{R}^{n_- \times n_-})$ in \eqref{obsprb} and $\hat{A}_{0}(z) \in \utriag(\mathbb{R}^{n_- \times n_-})$ in \eqref{obsfinendrb}, this leads to parallel stable transport equations \crefrange{hypsysot33}{hyprfref3ot3} driving the cascade of stable parabolic systems \crefrange{parasysoet3}{obsprb} and the cascade of stable transport equations \crefrange{hypsysot3}{obsfinendrb}, which is depicted in Fig. \ref{fig:obs2}.

Differentiating \eqref{obsfinaltraf} w.r.t. time and inserting \eqref{obset2} and \eqref{obset3} yields after the usual calculations 
the \emph{observer decoupling equations}%
\begin{subequations}\label{Ke:fso}
\begin{align}
(\bar{N}(z)\Lambda^-)_{\zeta}(\zeta) &= -\Theta \bar{N}_{zz} + \mu\o\bar{N} + \bar{R}_-(z,\zeta)\label{feqpdeo}\\
           \bar{N}_z(0,\zeta) &= - \tint_0^1\bar{F}_0(\bar{\zeta})\bar{N}(\bar{\zeta},\zeta)\d\bar{\zeta} + \bar{H}_{2,-}(\zeta)\label{feqrb1o}\\
             \bar{N}(1,\zeta) &= \bar{H}_{3,-}(\zeta)\label{feqrb2o}\\
                 \bar{N}(z,1)\Lambda^-(1) &= -M_w(z)(G_3^2)^{-1}\label{feqico}
\end{align}	 
\end{subequations}
defined on $(z,\zeta) \in [0,1]^2$ with $\bar{R}(z,\zeta) = -H_{1}^3(z)\delta(\zeta-z) - \bar{H}_1^4(z,\zeta)\sigma(z-\zeta) - \bar{H}_1^5(z,\zeta)  + M_w(z)(G_3^2)^{-1}(\bar{A}_0(\zeta) + G_3^2\bar{P}(1,1)\bar{H}_3^3(\zeta))$ (cf. appendix). The next lemma clarifies the solvability of the IBVP \eqref{Ke:fso}.
\begin{Lemma}[Observer decoupling equations]\label{lem:cfdecouplo}
 The observer decoupling equations \eqref{Ke:fso} have a unique mild solution $\bar{N}(\cdot,\zeta) \in (L_2(0,1))^{n_- \times n_-}$, $\zeta \in [0,1]$  if $\mu\o$ is such that \crefrange{parasysoet3}{obsprb} is exponentially stable.
\end{Lemma}
\begin{proof}
By making use of the tranformations $\tilde{N}(z,\zeta) = \Lambda^-(1-z)\bar{N}\t(\zeta,1-z)\Theta^{-1}(\zeta)$, 
the decoupling equations \eqref{Ke:fso} can be mapped into the form \eqref{Ke:fs}. This implies that \eqref{Ke:fso} have a unique mild solution in $L_2$.
\end{proof}
With the same calculations, the \emph{observer kernel equations}
\begin{subequations}\label{app:k2o}
	\begin{align}
	\Lambda^- \bar{Q}_{z} + (\bar{Q}(z)\Lambda^-)_{\zeta}(\zeta) &= 0, \; 0 < \zeta < z < 1\\
	\bar{Q}(1,\zeta) + \tint_\zeta^1\bar{\Gamma}_0(\bar{\zeta})\bar{Q}(\bar{\zeta},\zeta)\d\bar{\zeta} &= \bar{\Gamma}_0(\zeta) -\hat{A}_0(\zeta)\\ 
	\bar{Q}(z,z)\Lambda^-(z) - \Lambda^-(z)\bar{Q}(z,z) &= 0
	\end{align}	 
\end{subequations}
result for $\bar{Q}(z,\zeta)$ in \eqref{obsfinaltraf} with $\bar{\Gamma}_0(\zeta) = \bar{A}_{0,-}(\zeta) - G_3^2\bar{P}(1,1)\bar{H}_{3,-}(\zeta) - G_3^2\bar{N}_z(1,\zeta)$. These kernel equations have the same structure as \eqref{app:k2} so that they also have a piecewise $C^1$-solution $\bar{Q}(z,\zeta)$ (see \cite{Deu18a}). By making use of the corresponding reciprocity relation, the kernel $\bar{Q}_I(z,\zeta)$ of the inverse backstepping transformation related to \eqref{obsfinaltraf} can be found, while the decoupling transformation is directly solvable for $\tilde{e}_w$.

With this, the corresponding observer gains
\begin{subequations}\label{obsgain2}
\begin{align}
 \leftidx{^-}{\tilde{L}}(z) &= -\bar{Q}(z,0)\Lambda^-(0), \; \leftidx{^+}{\tilde{L}}(z) = 0\label{obsg2a}\\
 \tilde{L}_w(z) &= \tint_0^1\bar{N}(z,\bar{\zeta})\leftidx{^-}{\tilde{L}}(\bar{\zeta})\d\bar{\zeta} + \bar{H}_{1,-}^1(z) + \bar{N}(z,0)\Lambda^-(0)\label{obsg2b}
\end{align}	
\end{subequations}
can be determined.

The stability of the observer error dynamics is assessed in the next theorem, where the rate of the exponential decay can be assigned.
\begin{thm}[Observer]\label{zrstabobs}
Use the observer gains \eqref{obsgain1} and \eqref{obsgain2} for \eqref{obs}. Let $\bar{\mu}_{\text{max}}$ be the largest eigenvalue of \crefrange{parasysoet3}{obsprb} for $\mu\o = 0$ and choose the design parameter $\mu\o$  such that $\alpha\o = \bar{\mu}_{\text{max}} - \mu\o < 0$. 
Then, the observer error dynamics is exponentially stable in the norm $\|\cdot\| =  (\|\cdot\|_{\text{w}}^2 + \|\cdot\|_n^2)^{1/2}$, i.e., the observer error $e_{\text{o}}(t) = \{\col{e_w(z,t),e(z,t)}, z \in [0,1]\}$ satisfies
\begin{equation}
 \|e_{\text{o}}(t)\| \leq M_{\text{o}}\e^{(\alpha\o + c)t}\|e_{\text{o}}(0)\|, \quad  t \geq 0,
\end{equation}%
for an $M_{\text{o}} \geq 1$, all ICs $e_w(\cdot,0) \in (L_2(0,1))^{n_-}$,  piecewise differentiable ICs $e(z,0) \in \mathbb{R}^{n}$ and any $c > 0$ so that $\alpha\o + c < 0$. 	
\end{thm}	
Since the target systems \eqref{plantttf} and \eqref{obset3} share the same structure, the proof of Theorem \ref{zrstabobs} can be directly inferred from the proof of Theorem \ref{zrstab}.

\section{Observer-based compensator}\label{sec:obscomp}
For the design of an observer-based compensator, the following assumption is required.
\begin{ass}[Compensator design]\label{ass3}\hfill
 Assume that $\det H_3^1 \neq 0$, $\det G_3^2 \neq 0$, $H^2_3 = H_3^3 = 0$, $H_i = 0$, $i = 1,2$, $G_3^1 = G_3^3 = 0$ and $G_i = 0$, $i = 1,2$, in \eqref{plant}.
\end{ass} 
With this, the system \eqref{plant} is both in strict feedback and strict feedforward form. Hence, an observer-based compensator can be designed on the basis of the proposed backstepping approach by using the state estimates of the observer \eqref{obs} in the state feedback \eqref{sff1} and \eqref{sfb2}. The next theorem states the stability of the resulting closed-loop system. 
\begin{thm}[Observer-based compensator]\label{thm:sepprincf}
Consider the state feedback controller \eqref{sff1}, \eqref{sfb2} supplied with the estimates of the observer \eqref{obs} with the gains \eqref{obsgain1} and \eqref{obsgain2}. Assume that the state feedback controller and the observer have been designed such that $\alpha\c < 0$ and $\alpha\o < 0$ (see Theorem \ref{zrstab} and \ref{zrstabobs}). Then, the closed-loop system with the state $x_{\text{cl}}(t) = \operatorname{col}(e_w(t),e(t),\hat{w}(t),\hat{x}(t))$ is exponentially stable in the norm $\|\cdot\|\cl = (\|\cdot\|_{\text{w}}^2 +\linebreak \|\cdot\|_n^2 + \|\cdot\|_{\text{w}}^2 +\|\cdot\|_n^2)^{\frac{1}{2}}$ for all ICs $w(\cdot,0), \hat{w}(\cdot,0) \in (L_2(0,1))^{n_-}$ and  piecewise differentiable ICs $x(z,0), \hat{x}(z,0) \in \mathbb{R}^{n}$.
\end{thm}
\begin{proof}
Consider the observer \eqref{obs} in view of the Assumptions \ref{ass3} and successively apply the transformations \eqref{bcoord} and \eqref{firstdecouplp} as well as use the feedback \eqref{sff1}, \eqref{sfb2} with the state estimates. After homogenizing the BCs, this yields a system of the form
\begin{subequations}\label{plantttfc}
\begin{align}
	\dot{\check{\varepsilon}}_- &= \Lambda^-(z)\check{\varepsilon}'_- + \check{A}_0(z)\check{\varepsilon}_-(0) + \rho_1(z,\varepsilon_w(0),\bar{e}_-)\label{hypsysttfc}\\
	\check{\varepsilon}_-(1) &= 0\label{hyprfref3ttcf}\\
	\dot{\check{w}} &= \Theta(z)\check{w}'' - \mu\c\check{w} + \tilde{F}_{0}(z)\check{w}(0)\label{tdiv}\\
	                &\hspace{-0.55cm}  + \rho_2(z,\tilde{e}_+(1),\tilde{e}_+,\varepsilon_w,\varepsilon_w(0),\dot{\varepsilon}_w(0), \varepsilon_w(1),\varepsilon_w'(1),\bar{e}_-(0),\bar{e}_-)\nonumber \\
	\check{w}'(0) &= 0, \; \check{w}(1) = H_3^1\check{\varepsilon}_-(0)\label{parasys3ttfc}\\
	\dot{\check{x}}_+ &= \Lambda^+(z)\check{x}'_+ + \leftidx{^+}{\tilde{A}}_0(z)\check{\varepsilon}_-(0) + \rho_3(z,\check{w},\varepsilon_w(0),\bar{e}_-) \label{hypsysttf2c}\\
	\check{x}_+(0) &= Q_+\check{\varepsilon}_-(0) + \rho_4(\check{w},\varepsilon_w(0),\bar{e}_-)\label{hypendc}	              
\end{align}
\end{subequations}
with some functions $\rho_i$, $i = 1,\ldots,4$. First, note that the latter functions are well-defined, because their arguments give rise to relatively bounded operators, which is verified in \cite{Deu20}. Furthermore, also the time derivative $\dot{\varepsilon}_w(0)$ in \eqref{tdiv} exists, which is due to the smooth solution of the parabolic $\varepsilon_{w}$-subsystem. In particular, the corresponding system operator is the generator of an analytic $C_0$-semigroup (see the analysis of the observer in \cite{Deu19} for details).
With this, the excitation via the $\rho_i$-functions decays exponentially for $t \geq 0$ due to Theorem \ref{zrstabobs}. Since \eqref{plantttfc} coincides with the target system \eqref{plantttf} for the state feedback design except for the exciting terms, from the proof of Theorem \ref{zrstab}, it follows that the solution decays exponentially in the norm $\|\cdot\|$. Hence, the corresponding closed-loop system  represented by the cascade composed of the observer error dynamics \eqref{obse} and the observer \eqref{plantttfc} is  exponential stable in the norm $\|\cdot\|\cl$.
Then, the same is true in the original coordinates, because of the bounded invertibility of the used transformations. 
\end{proof}

\begin{figure}[t]\centering	
	\hspace*{-0.43cm} 
	\begin{tabular}{c@{\hskip 0.01in}c}
		\includegraphics[width=0.5\columnwidth]{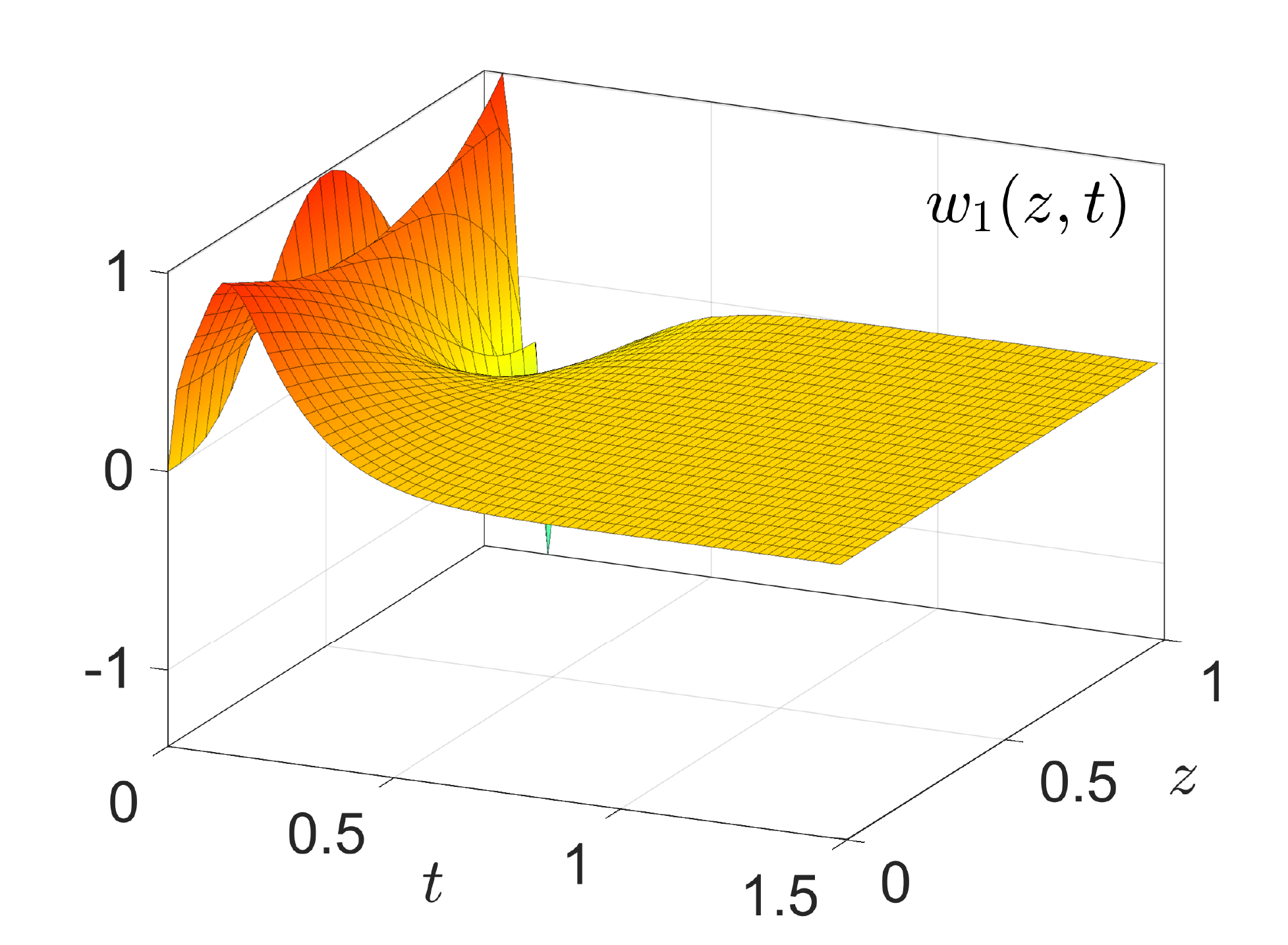}  & 	\includegraphics[width=0.5\columnwidth]{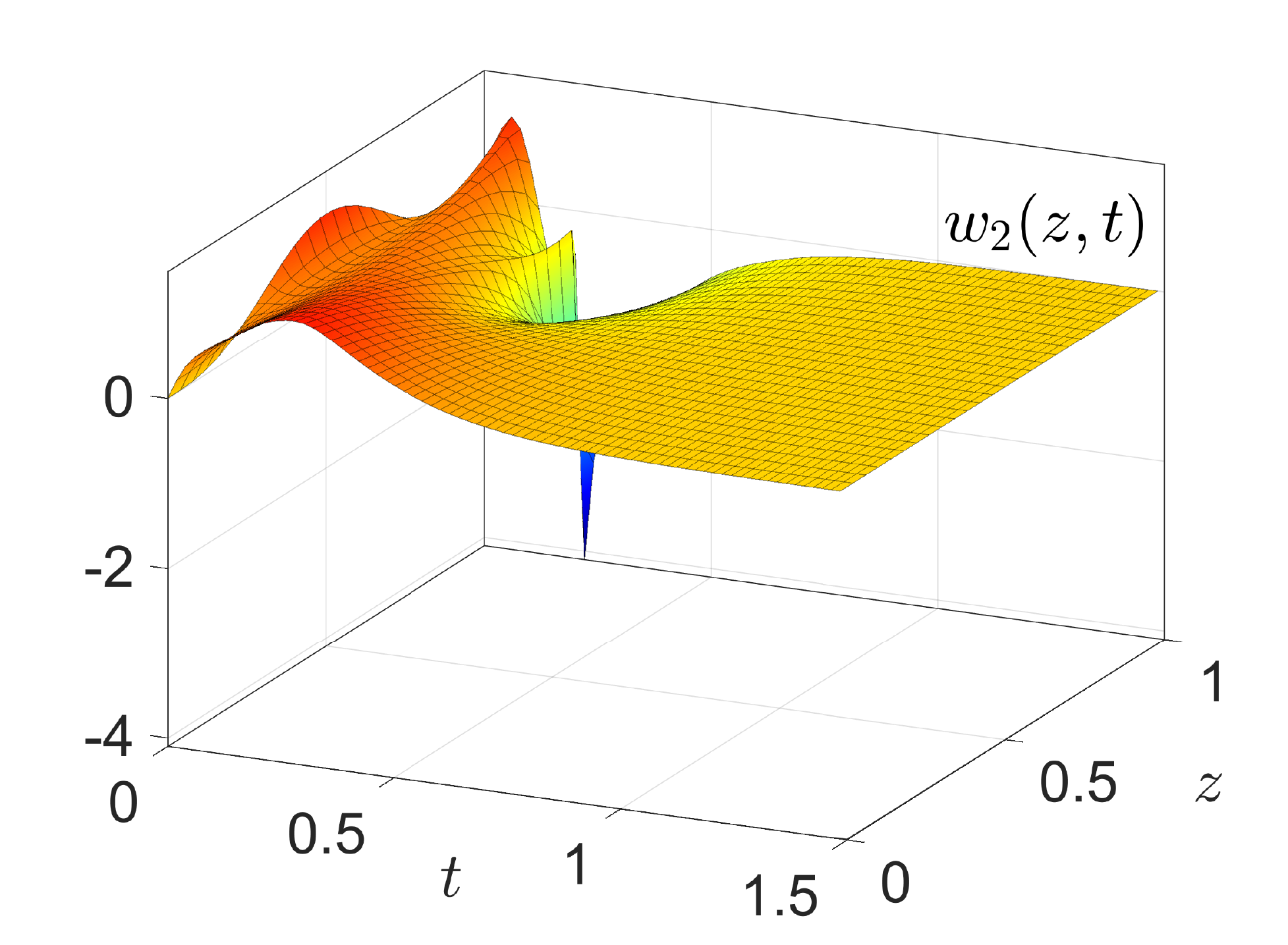}\\
		\includegraphics[width=0.5\columnwidth]{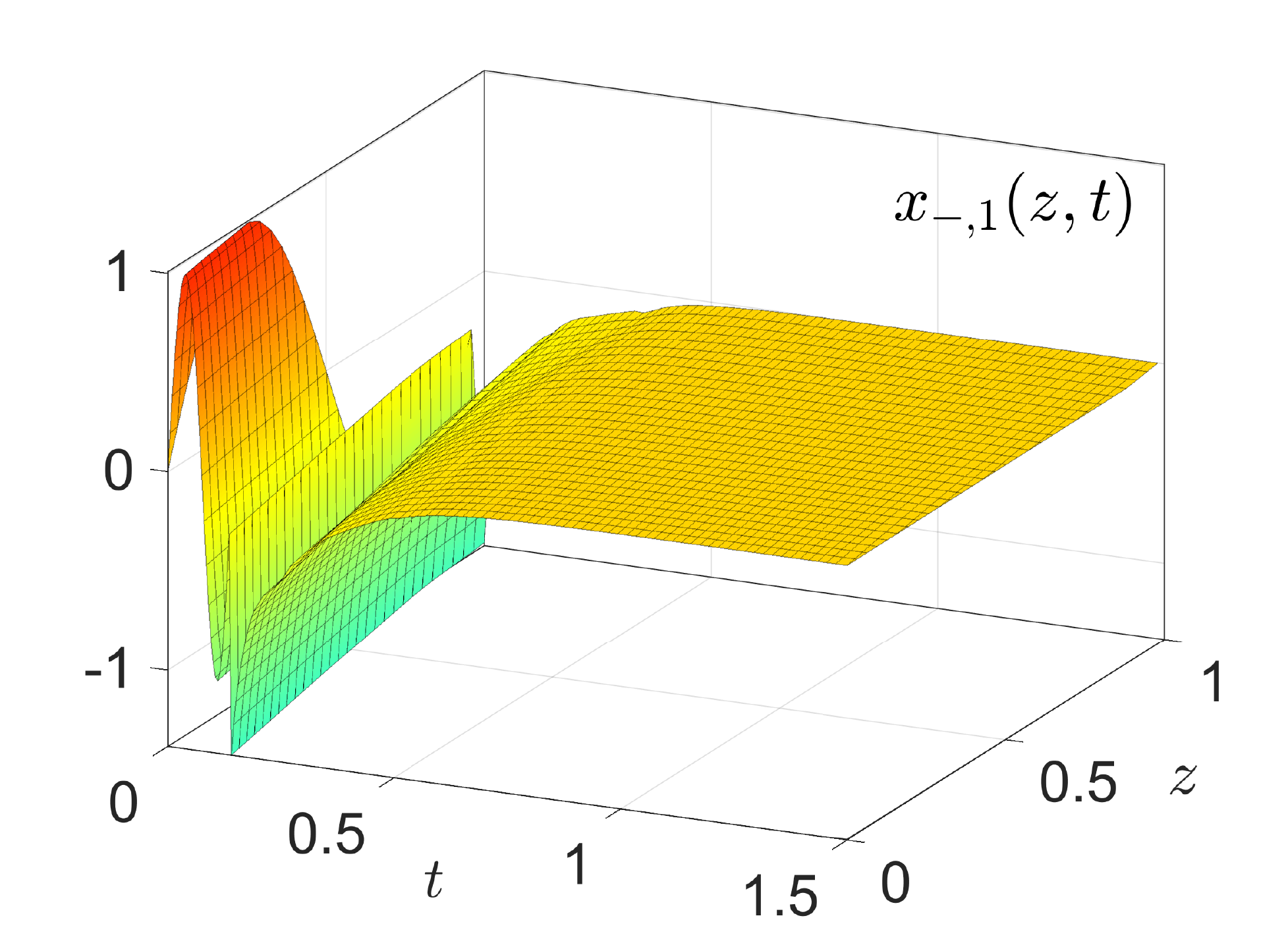}	&	\includegraphics[width=0.5\columnwidth]{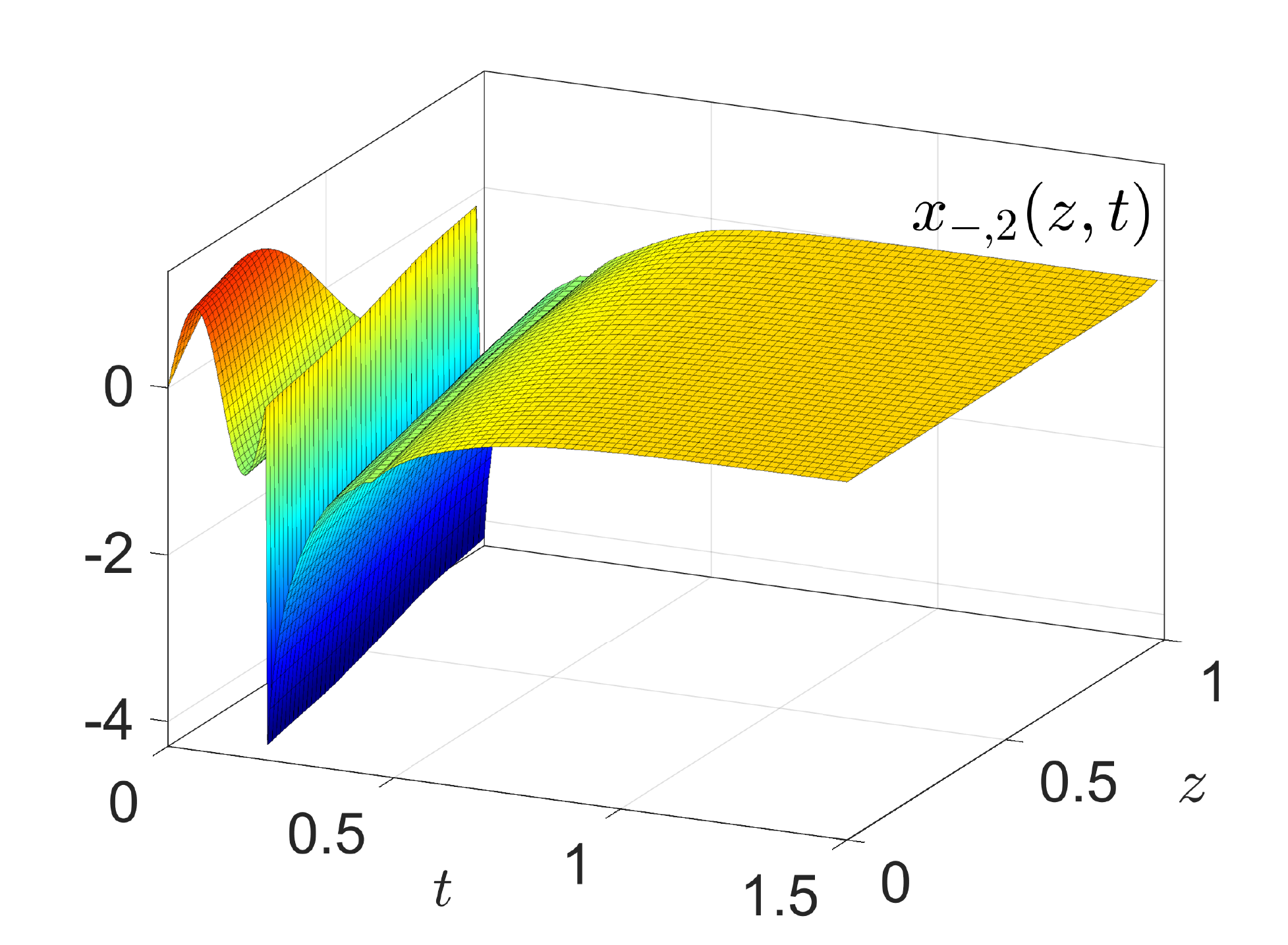}\\
		\includegraphics[width=0.5\columnwidth]{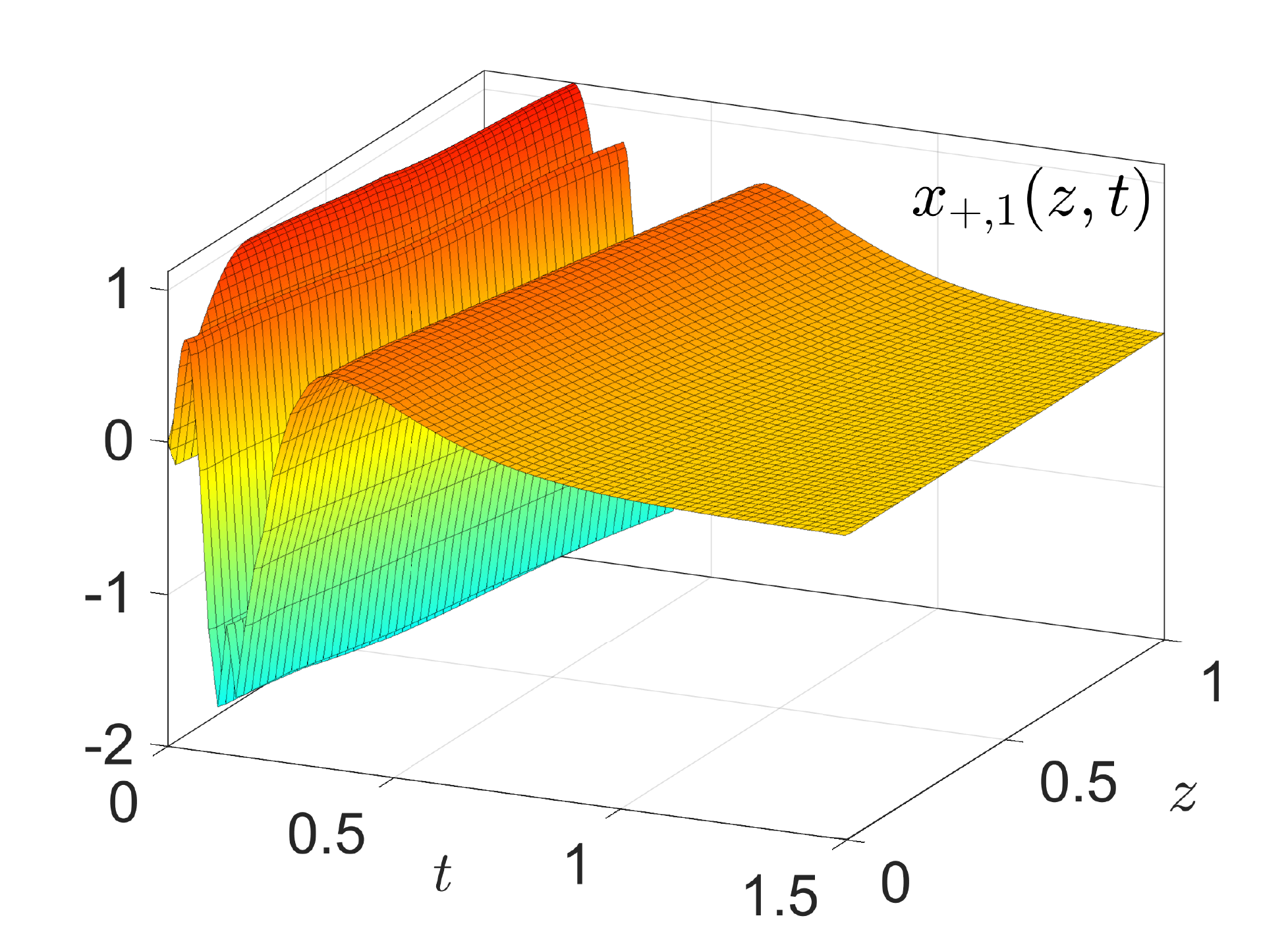} & 	\includegraphics[width=0.5\columnwidth]{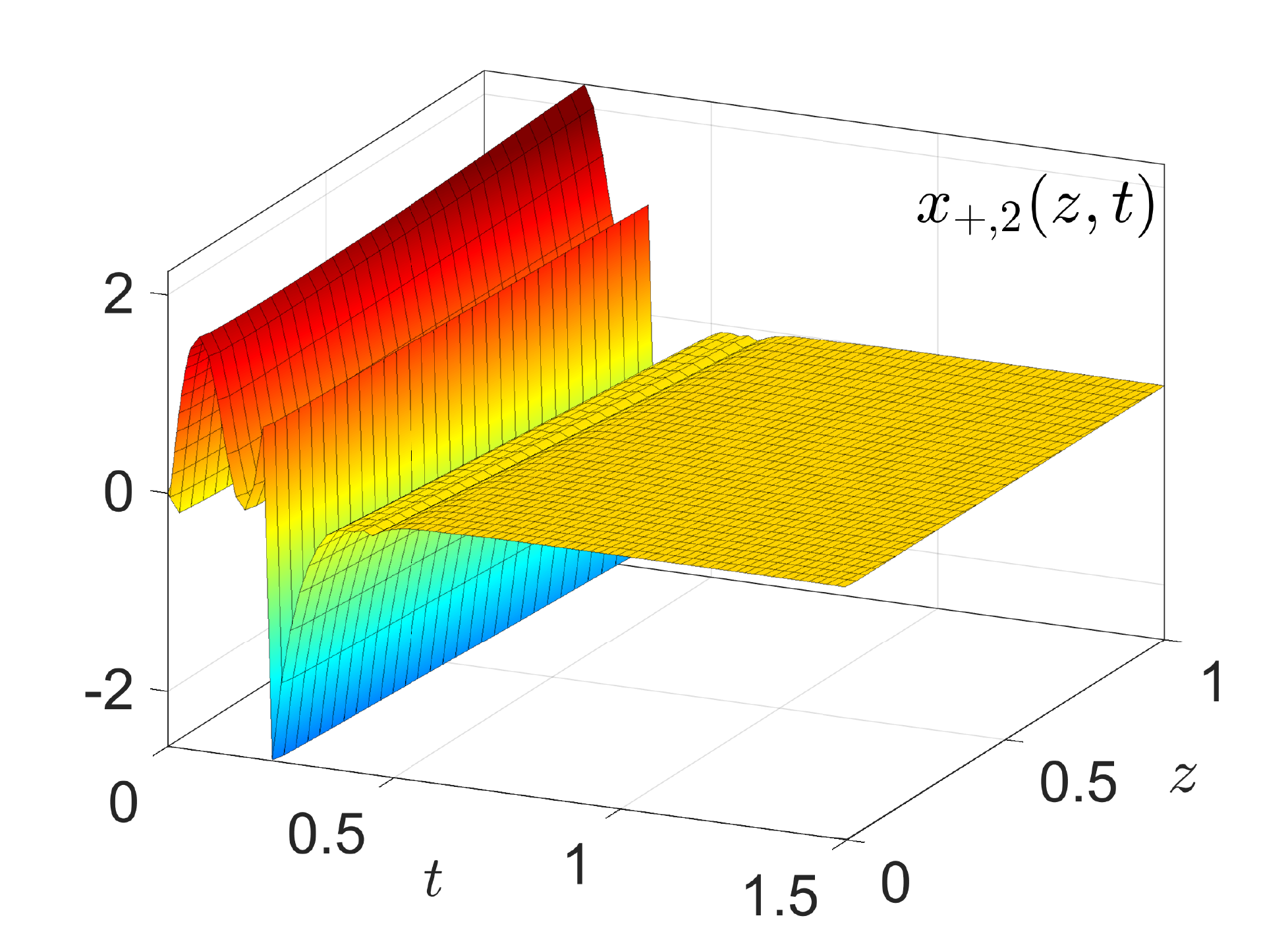}
	\end{tabular}	
	%
	\caption{Simulation results for the state feedback loop design. The first row depicts the closed-loop states of the parabolic subsystem, while the remaining rows show the closed-loop states of the hyperbolic subsystem.}\label{fig:ex1}%
\end{figure}
\section{Examples}\label{sec:ex}
\textbf{Problem Setup.} 
In order to illustrate the results of the paper, consider the unstable hyperbolic-parabolic PDE-PDE system \eqref{plant} with the matrix $\Theta(z) = \diag{2+\frac{3}{5}\sin(2 \pi z),1+z^2 \sin(2 \pi z)}$ of diffusion coefficients, the matrix $\Lambda(z) = \diag{6+2z,4+z,-3.5 - \frac{z}{2},-5.75 - \frac{3z}{2}}$ of transport velocities and 
\begin{subequations}
	\begin{align}
	F(z) &= \begin{bmatrix}
      	     0                        & \frac{3}{4} \text{e}^{-z} \\
            \frac{3}{4} \text{e}^{-z} & \frac{11}{5} \text{sin}(2 \pi z)
	\end{bmatrix},\\
	A(z) &= \begin{bmatrix}
	          0                           & 0.3 \, \text{e}^\frac{z}{3} & 0                          & -\frac{1}{6} - \frac{z}{3} \\
	         -0.3 \, \text{e}^\frac{z}{3} & 0                           & -\frac{1}{4} - \frac{z}{2} & 0 \\
	                                    0 & \frac{1}{4} + \frac{z}{2}   & 0                          & 0.2 \, \text{e}^{-\frac{z}{2}} \\
	            \frac{1}{6} + \frac{z}{3} & 0                           & -\frac{1}{5} \, \text{e}^{-\frac{z}{2}} & 0
	\end{bmatrix},
\end{align}	
\end{subequations}%
$Q_0 = Q_+ = I_2$, $Q_- = -2I_2$ characterizing the parabolic and hyperbolic subsystems. Their bilateral interconnection is described by the coupling matrix $H_3^1 = I_2$ for the parabolic subsystem as well as 
$\vect{G_1^1(z)} = \col{0, \text{cos}(2 \pi z),-\text{sin}(2 \pi z),\linebreak  \text{e}^{z+1},(1-z)z,0,0,0,0}$, 
$\vect{G^3_1(z)} = \col{(1-z)z,0,0, \text{sin}^3(2 \pi z),\linebreak  0,\frac{1}{2} \text{sin}(2 \pi z),2 (1-z)^3,0}$  
$\vect{G^4_1(z,\zeta)} = \col{0,\text{e}^\frac{z}{3} \text{cos}(2 \pi \zeta),\linebreak  0,0,\text{cos}(2 \pi z) \text{e}^\frac{\zeta}{2},\frac{1}{2} \text{e}^\frac{z}{3} \text{sin}(2 \pi \zeta), 2 \text{e}^\frac{\zeta}{2},\text{log}(1+z)\text{e}^\frac{\zeta}{2}}$, 
$\vect{G_2^1} = \col{\linebreak  -1,0,1,1}$,
$G_3^2 = I_2$ and $\vect{G_3^3(z)} = \col{1,0,z,2+z^2}$ for the hyperbolic subsystem. Therein, $\operatorname{vec}(\cdot)$ is the usual vectorization operator. All other matrices appearing in \eqref{HGmat} are zero. Hence, the system is in strict feedback form. In what follows, firstly, a backstepping state feedback controller is determined to stabilize this system. 
For the compensator design, a second unstable example that is both in strict feedback and strict feedforward form is considered. The latter follows from the previous example by setting all coupling matrices to zero that violate Definition \ref{ass2}. 

\textbf{State Feedback Controller.} 
For the design of the state feedback controller, 
the kernel equations \eqref{app:k1} with $\mu_{\text{c}} = 3$ are solved in MATLAB   by implementing the method of successive approximations presented in \cite{Deu17,Hu15a,Hu19}. With this, a solution of the decoupling equations \eqref{Ke:fs} is determined by making use of the MATLAB solver \texttt{pdepe} with $71$ grid points. Finally, the kernel equations \eqref{app:k2} are also solved using the method of successive approximations, which yields the gains of the state feedback controller with help of  \eqref{sff1} and \eqref{sfb2}. The simulation of the closed-loop system uses the IC $w(z,0) = \col{1,1}(\frac{3}{4}\text{sin}(\pi z)+\frac{1}{4}\text{cos}(3 \pi z + \frac{\pi}{2}))$ for the parabolic subsystem and $x(z,0) = 
\col{1,1,1,1,1}\text{sin}(2 \pi z)$	for the hyperbolic subsystem. Note that these ICs are consistent both w.r.t. the BCs of the corresponding subsystems as well as w.r.t. the coupling at the boundary $z=1$ in \eqref{parasys3}. The parabolic PDE is simulated using a FEM  and the hyperbolic PDE is discretized using a FDM ensuring numerical stability. This yields the plots of the resulting state feedback loop shown in Fig.  \ref{fig:ex1} 
according to Theorem \ref{zrstab}. 
\begin{figure}[t]\centering
	\begin{tabular}{c@{\hskip 0.05in}c}
    	\includegraphics[width=0.45\columnwidth]{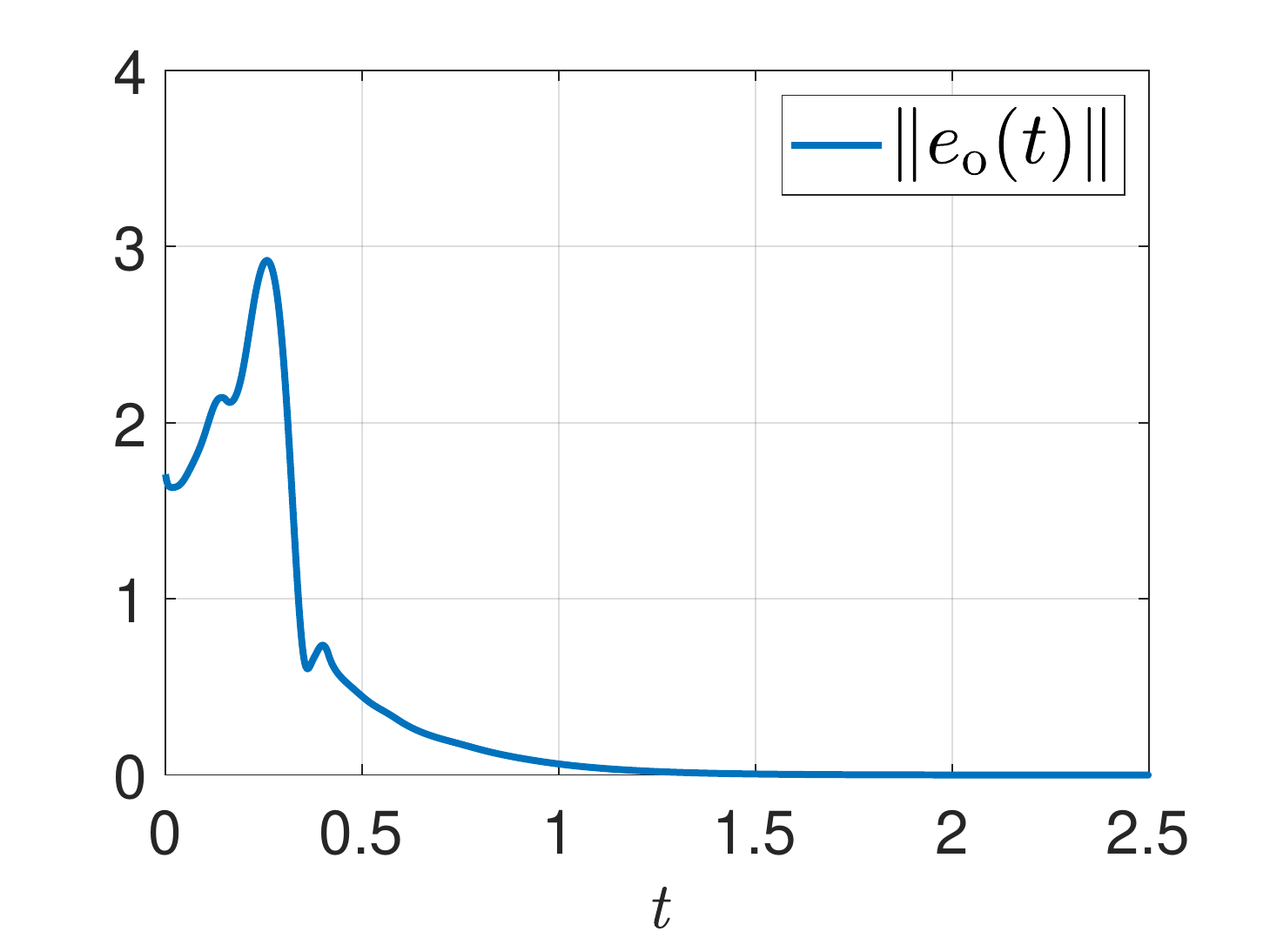} &
		\includegraphics[width=0.45\columnwidth]{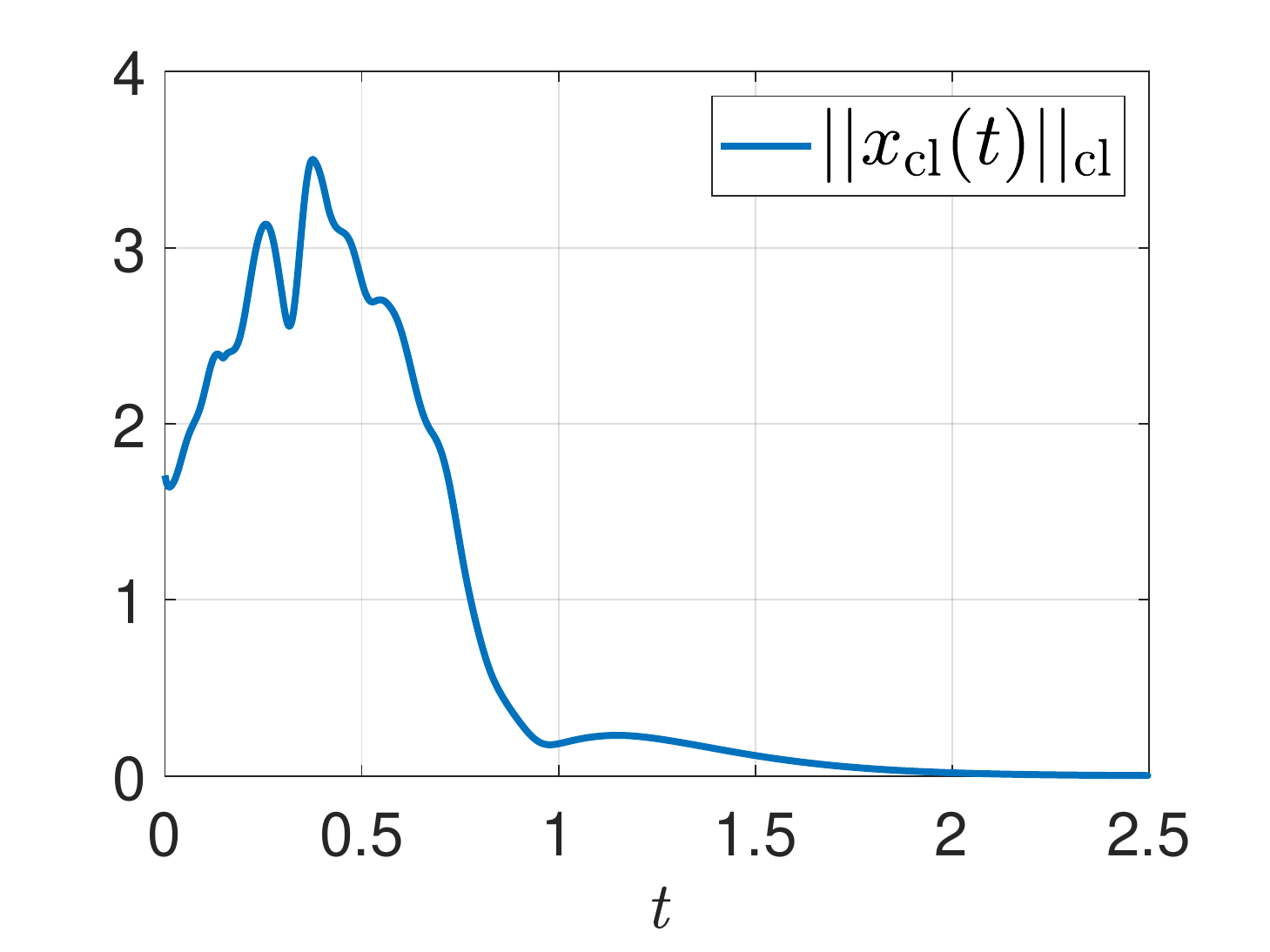}	
	\end{tabular}
	\caption{Time evolution of the $L_2$-norms of the observer error  (left) and the closed-loop system with observer-based compensator (right). 
    \label{fig:ex2}}
\end{figure}%
\textbf{Observer-based Compensator.} 
The second example considers the design of an observer-based compensator. For this, a state feedback controller is determined along the lines  of the previous example 
with reduced couplings. Then, a backstepping observer is designed to implement the corresponding state feedback. For this, the observer error dynamics is mapped into backstepping coordinates with the design parameter $\mu_{\text{o}} = 6$. The kernel equations \eqref{app:k1o} are solved using the method of successive approximations in MATLAB (see \cite{Deu17,Hu15a,Hu19}). Subsequently, a solution of the decoupling equations \eqref{Ke:fso} is determined with the MATLAB solver \texttt{pdepe} for $71$ grid points. The observer design is completed by solving the kernel equations \eqref{app:k2o} with the method of successive approximations. Different from the state feedback controller, however, the calculation of the observer gains has to be done backwards. In particular, one has to follow the following sequence starting from \eqref{obsg2a}: $\tilde{L}(z) \xrightarrow{\eqref{obsg2b}} \tilde{L}_w(z) \xrightarrow{\text{\crefrange{obsgain1a}{obsgain1b}}} L_w(z), L(z)$. The remaining observer gains  are computed directly. The $L_2$-norm of the observer error is depicted in Fig. \ref{fig:ex2} on the left for a zero observer IC and assuming the plant IC  according to the first example, which validates a convergent observer. By combining the state feedback controller with the observer, the plot shown in Fig. \ref{fig:ex2} on the right is obtained. 
This result verifies closed-loop stability with the observer-based compensator. 

\section{Concluding Remarks}
It is interesting to also apply the proposed design procedure to parabolic-hyperbolic, parabolic-parabolic or hyperbolic-hyperbolic PDE-PDE systems. Then, only different decoupling equations appear in the design, which have to be analyzed for well-posedness. 
A salient feature of the proposed multi-step approach is its \emph{scalability property}. In particular, for a next neighbor coupling between more than two PDEs, the design does not become more involved, as only the number of the same kernel and decoupling equations increases. This is crucial for considering networks of PDEs in arbitrary interconnection topologies, which will be investigated in future work. Furthermore, it is of interest to ensure  finite-time stability for both the parabolic and hyperbolic subsystems, which requires to consider a time-varying target system and thus time-varying backstepping transformations. 
Another research avenue are fully coupled hyperbolic-parabolic system, which require a detailed controllability and obervability analysis and new techniques for the successive transformation approach.

\appendix\label{app}

\textbf{Operators $\pmb{\tilde{\mathcal{G}}_i[\tilde{w}],\; i = 1,2,3}$.} 
Consider first the operators $\tilde{\mathcal{G}}_i[\tilde{w}] =  G_i^1\tilde{w}(0) + G_i^2\tilde{w}'(1) + G_i^2P_I(1,1)\tilde{w}(1) + \tint_0^1\tilde{G}^3_i(\zeta)\tilde{w}(\zeta)\d\zeta$, $i = 2,3$, with 
$\tilde{G}^3_2(\zeta) = Q_+P_I(1,\zeta) +G_2^2P_{I,z}(1,\zeta) + G_2^3(\zeta) + \tint_{\zeta}^1(G_2^3(\bar{\zeta})P_I(\bar{\zeta},\zeta)\d\zeta$ and $\tilde{G}^3_3(\zeta) = G_3^2P_{I,z}(1,\zeta) + G_3^3(\zeta) + \tint_{\zeta}^1G_3^3(\bar{\zeta})P_I(\bar{\zeta},\zeta)\d\zeta$.
 Then, the operator
$\tilde{\mathcal{G}}_1[\tilde{w}](z) =  \tint_0^1\tilde{G}_1(z,\zeta)\tilde{w}(\zeta)\d\zeta$ is given by 
$\tilde{G}_1(z,\zeta) = \tilde{G}^1_{1}(z)\delta(\zeta) + \tilde{G}^2_{1}(z)\delta'(\zeta-1) + \tilde{G}_1^2(z)P_I(1,1)\delta(\zeta-1) + G^3_1(z)\delta(\zeta-z) +  \tilde{G}^4_{1}(z,\zeta)\sigma(\zeta-z) + \tilde{G}^5_{1}(z,\zeta)$ with 
$\tilde{G}_1^i(z) = \mathcal{T}_x[G_1^i](z) + K(z,0)\Lambda_+(0)G_2^i$, $i = 1,2$,
$\tilde{G}_1^4(z,\zeta) = \check{G}_1^4(z,\zeta)\linebreak - K(z,\zeta)G_1^3(\zeta) - \tint_{\zeta}^zK(z,\bar{\zeta})\check{G}_1^4(\bar{\zeta},\zeta)\d\bar{\zeta}$, 
$\tilde{G}_1^5(z,\zeta) = \mathcal{T}_x[\check{G}^5_1(\cdot,\zeta)](z) + K(z,0)\Lambda_+(0)\tilde{G}_2^3(\zeta)$, where  
$\check{G}_1^4(z,\zeta) = G_{1}^3(z)P_I(z,\zeta) + G_1^4(z,\zeta) + \tint_{\zeta}^zG_1^4(z,\bar{\zeta})P_I(\bar{\zeta},\zeta)\d\bar{\zeta}$ and 
$\check{G}_1^5(z,\zeta) = G_{1}^2(z)P_{I,z}(1,\zeta) + G_1^5(z,\zeta) + \tint_{\zeta}^1G_1^5(z,\bar{\zeta})P_I(\bar{\zeta},\zeta)\d\bar{\zeta}$.

\textbf{Operators $\pmb{\bar{\mathcal{H}}_i[\tilde{e}],\; i = 1,2,3}$.} 
The operators $\bar{\mathcal{H}}_i[\tilde{e}] =  H_i^2\tilde{e}_+(1)\linebreak + \tint_0^1\bar{H}^3_i(\zeta)\tilde{e}(\zeta)\d\zeta$, $i = 2,3$, are given by
$\bar{H}^3_i(\zeta) = H_i^3(\zeta) - \tint_{\zeta}^1H_i^3(\bar{\zeta})\bar{K}(\bar{\zeta},\zeta)\d\bar{\zeta} - H^2_i\leftidx{^+}{\bar{K}}(1,\zeta)$. Then, $\bar{\mathcal{H}}_1[\tilde{e}](z) = \linebreak\tint_0^1\bar{H}_1(z,\zeta)\tilde{e}(\zeta)\d\zeta$ with
$\bar{H}_1(z,\zeta) = \bar{H}_1^1(z)E_-\t\delta(\zeta) + \bar{H}_1^2(z)E_+\t\delta(\zeta-1) + H_1^3(z)\delta(\zeta-z) + \bar{H}_1^4(z,\zeta)\sigma(\zeta-z) + \bar{H}_1^5(z,\zeta)$,\linebreak in which
$\bar{H}_1^1(z) = \bar{\mathcal{T}}_w[H_1^1](z)$, $\bar{H}_1^2(z) = \bar{\mathcal{T}}_w[H_1^2](z) - \bar{\mathcal{T}}_w[(\bar{P}\Theta)_{\zeta}(\cdot,1)](z)H_3^2$, $\tilde{H}_1^4(z,\zeta) = H_1 ^4(z,\zeta) - \tint_{\zeta}^zH_1^4(z,\bar{\zeta)}\linebreak\cdot\bar{K}(\bar{\zeta},\zeta)\d\bar{\zeta} - H_1^3(z)\bar{K}(z,\zeta)$, $\bar{H}_1^4(z,\zeta) = \bar{\mathcal{T}}_w[\tilde{H}_1^4(\cdot,\zeta)](z) - \bar{P}_I(z,\zeta)H_1^3(\zeta) - \tint_{\zeta}^1\bar{P}_I(z,\bar{\zeta})\tilde{H}_1 ^4(\bar{\zeta},\zeta)\d\bar{\zeta}$, $\tilde{H}_1^5(z,\zeta) = H_1 ^5(z,\zeta)\linebreak - \tint_{\zeta}^1H_1^5(z,\bar{\zeta)}\bar{K}(\bar{\zeta},\zeta)\d\bar{\zeta}- H_1^2(z)\leftidx{^+}{\bar{K}}(1,\zeta)$ and $\bar{H}_1^5(z,\zeta) = \bar{\mathcal{T}}_w[\tilde{H}_1^5(\cdot,\zeta)](z) + \bar{P}_I(z,\zeta)H_1^3(\zeta) + \tint_{\zeta}^1\bar{P}_I(z,\bar{\zeta})\tilde{H}_1^4(\bar{\zeta},\zeta)\d\bar{\zeta} - \bar{\mathcal{T}}_w[\bar{P}_\zeta(\cdot,1)\Theta(1)](z)\bar{H}_3^3(\zeta)$.

\bibliographystyle{IEEEtranS}

\end{document}